\documentclass[12pt]{article}

\usepackage{scicite}

\usepackage{times}
\usepackage{graphicx}
\usepackage{multirow}
\usepackage{graphics}
\usepackage{csvsimple,longtable,booktabs}
\usepackage{pdflscape}
\usepackage{tabularx}
\usepackage{float}
\usepackage{xr}

\usepackage{color}

\usepackage{times}
\usepackage{graphicx}
\usepackage{multirow}
\usepackage{graphics}
\usepackage{csvsimple,longtable,booktabs}
\usepackage{pdflscape}
\usepackage{tabularx}
\usepackage{float}
\usepackage{xr}

\newenvironment{sciabstract}{%
\begin{quote} \bf}
{\end{quote}}

\newcommand{\beginsupplement}{%
        \setcounter{table}{0}
        \renewcommand{\thetable}{S\arabic{table}}%
        \setcounter{figure}{0}
        \renewcommand{\thefigure}{S\arabic{figure}}%
     }

\newcounter{lastnote}

\title{Structure, Function, and Control of the Musculoskeletal Network}

\author
{Andrew C. Murphy$^{1,2}$, Sarah F. Muldoon$^{1,3}$, David Baker$^{1,4}$, Adam Lastowka$^5$, \\
Brittany Bennett$^{5,6}$, Muzhi Yang$^{1,7}$, \& Danielle S. Bassett$^{1,4,8\ast}$\\
\\
\normalsize{$^{1}$ Department of Bioengineering, University of Pennsylvania, Philadelphia, PA 19104, USA}\\
\normalsize{$^{2}$ Perelman School of Medicine, University of Pennsylvania, Philadelphia, PA 19104, USA}\\
\normalsize{$^{3}$ Department of Mathematics, University of Buffalo, Buffalo, NY USA}\\
\normalsize{$^{4}$ Department of Electrical and Systems Engineering,}\\
\normalsize{University of Pennsylvania, Philadelphia, PA 19104 USA}\\
\normalsize{$^{5}$ Open Connections, Newtown Square, PA 19073 USA}\\
\normalsize{$^{6}$ Philadelphia Academy of Fine Arts, Philadelphia, PA 19102}\\
\normalsize{$^{7}$ Applied Mathematical and Computational Science Graduate Group,}\\
\normalsize{University of Pennsylvania, Philadelphia, PA 19104, USA}\\
\\
\normalsize{$^\ast$To whom correspondence should be addressed; E-mail:  dsb@seas.upenn.edu.}
}

\date{}

\begin{document} 


\baselineskip24pt


\maketitle


\begin{sciabstract}
\newpage
The human body is a complex organism whose gross mechanical properties are enabled by an interconnected musculoskeletal network controlled by the nervous system. The nature of musculoskeletal interconnection facilitates stability, voluntary movement, and robustness to injury. However, a fundamental understanding of this network and its control by neural systems has remained elusive. Here we utilize medical databases and mathematical modeling to reveal the organizational structure, predicted function, and neural control of the musculoskeletal system. We construct a whole-body musculoskeletal network in which single muscles connect to multiple bones via both origin and insertion points. We demonstrate that a muscle's role in this network predicts susceptibility of surrounding components to secondary injury. Finally, we illustrate that sets of muscles cluster into network communities that mimic the organization of motor cortex control modules. This novel formalism for describing interactions between the muscular and skeletal systems serves as a foundation to develop and test therapeutic responses to injury, inspiring future advances in clinical treatments. 
\end{sciabstract}
\newpage


The interconnected nature of the human body has long been the subject of both scientific inquiry and superstitious beliefs. From the ancient humors linking heart, liver, spleen, and brain with courage, calm, and hope\cite{Bynum1997}, to the modern appreciation of the gut-brain connection\cite{Wang2008}, humans tend to search for interconnections between disparate parts of the body to explain complex phenomena. Yet, a tension remains between this basic conceptualization of the human body and the reductionism implicit in modern science\cite{Craver2009}. An understanding of the entire system is often relegated to a futuristic world, while individual experiments fine-tune our understanding of minute component parts.

The human musculoskeletal system is no exception to this dichotomy.  While medical practice focuses in hand, foot, or ankle, clinicians know that injuries to a single part of the musculoskeletal system necessarily impinge on the workings of other (even remotely distant) parts\cite{neumann2013}. An injury to an ankle can alter gait patterns, leading to chronic back pain; an injury to a shoulder can alter posture, causing radiating neck discomfort. Understanding the fundamental relationships between focal structure and potential distant interactions requires a holistic approach. Here, we will detail such an approach that is able to account for structure, function, and control of the musculoskeletal system.

We specifically apply this framework to the problem of rehabilitation following injury to either skeletal muscle or cerebral cortex. Direct injury to a muscle or associated tendon or ligament affects other muscles via compensatory mechanisms of the body \cite{Colne2006}. Similarly, loss of use of a particular muscle or muscle group from direct cortical insult can result in compensatory use of alternate muscles \cite{Lum2009,Roby2003}. How the interconnections of the musculoskeletal system are structured and how they function, directly constrains how injury to a certain muscle will affect the musculoskeletal system as a whole. Understanding these interconnections could provide much needed insight into which muscles are most at risk for secondary injury due to compensatory changes resulting from focal injury, thereby informing more comprehensive approaches to rehabilitation following muscle injury. Additionally, an understanding of how the cortex maps onto not only single muscles, but also muscle subnetworks, could inform future empirical studies of the relationships between focal injuries (including stroke) to motor cortex and risk for secondary injury.

\section*{Results}

\subsection*{Structure of the musculoskeletal network}

\begin{figure}
\centerline{\includegraphics{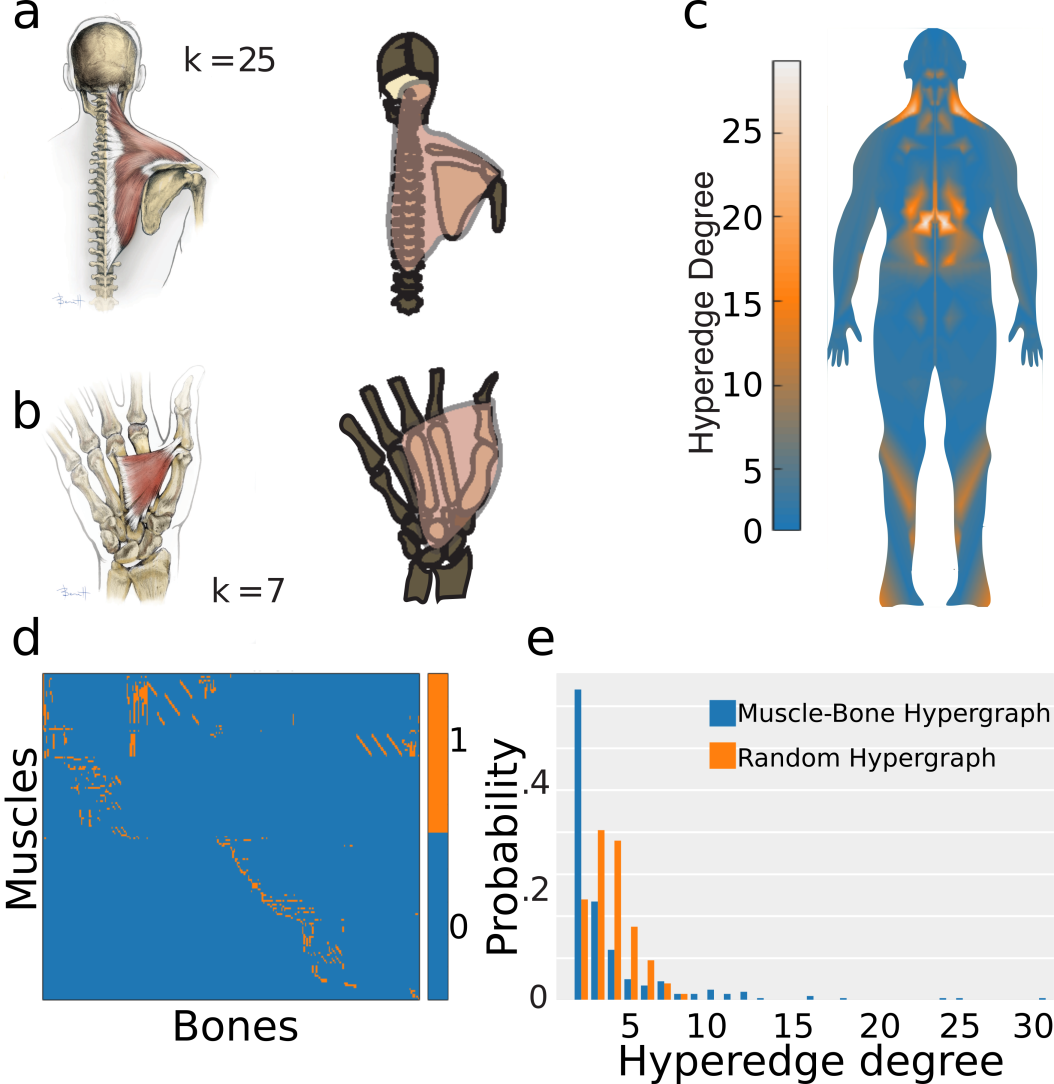}}
\caption{\linespread{1.5}\selectfont{} \textbf{Hypergraph structure.}  (a) Left: Anatomical drawing highlighting the trapezius.  Right:  Transformation of the trapezius into a hyperdege (red; degree $k=25$), linking 25 nodes (bones) across the head, shoulder, and spine.  (b)  Adductor pollicis muscle linking 7 bones in the hand.  (c) Spatial projection of the hyperedge degree distribution onto the human body. High-degree hyperedges are most heavily concentrated at the core.  (d) The musculoskeletal network displayed as a bipartite matrix (1 = connected, 0 otherwise).  (e) The hyperedge degree distributions for the musculoskeletal hypergraph, which is significantly different than that expected in a random hypergraph. \label{fig1}}
\end{figure}

To examine the structural interconnections of the human musculoskeletal system, we used a hypergraph approach. Drawing from recent advances in network science\cite{Newman2010}, we examined the musculoskeletal system as a network in which bones (network nodes) are connected to one another by muscles (network hyperedges). A hyperedge is an object that connects multiple nodes; muscles link multiple bones via origin and insertion points. The degree, $k$, of a hyperedge is equal to the number of nodes it connects; thus, the degree of a muscle is the number of bones it contacts. For instance, the \emph{trapezius} is a high degree hyperedge that links 25 bones throughout the shoulder blade and spine; conversely, the \emph{adductor pollicis} is a low degree hyperedge that links 7 bones in the hand (Fig.~\ref{fig1}a--b). A collection of hyperedges (muscles) that share nodes (bones) is referred to as a hypergraph: a graph $\mathcal{H} = (V, E)$ with $N$ nodes and $M$ hyperedges, where $V=\{v_1,\cdots,v_N\}$ is the set of nodes and $E=\{e_1,\cdots, e_M\}$ is the set of hyperedges (Fig.~\ref{fig1}d).  

The representation of the human musculoskeletal system as a hypergraph facilitates a quantitative assessment of its structure. We observed that the distribution of hyperedge degree is heavy-tailed: most muscles link 2 bones and a few muscles link many bones. The skew of the degree distribution differs significantly from that of random networks (two-sample Kolmogorov-Smirnov test, $p \ll .001$, see Methods)\cite{Newman2010}, indicating the presence of muscles of unexpectedly low and high degree (Fig.~\ref{fig1}e).

\subsection*{Function of the musculoskeletal network}

\begin{figure}
\centerline{\includegraphics[]{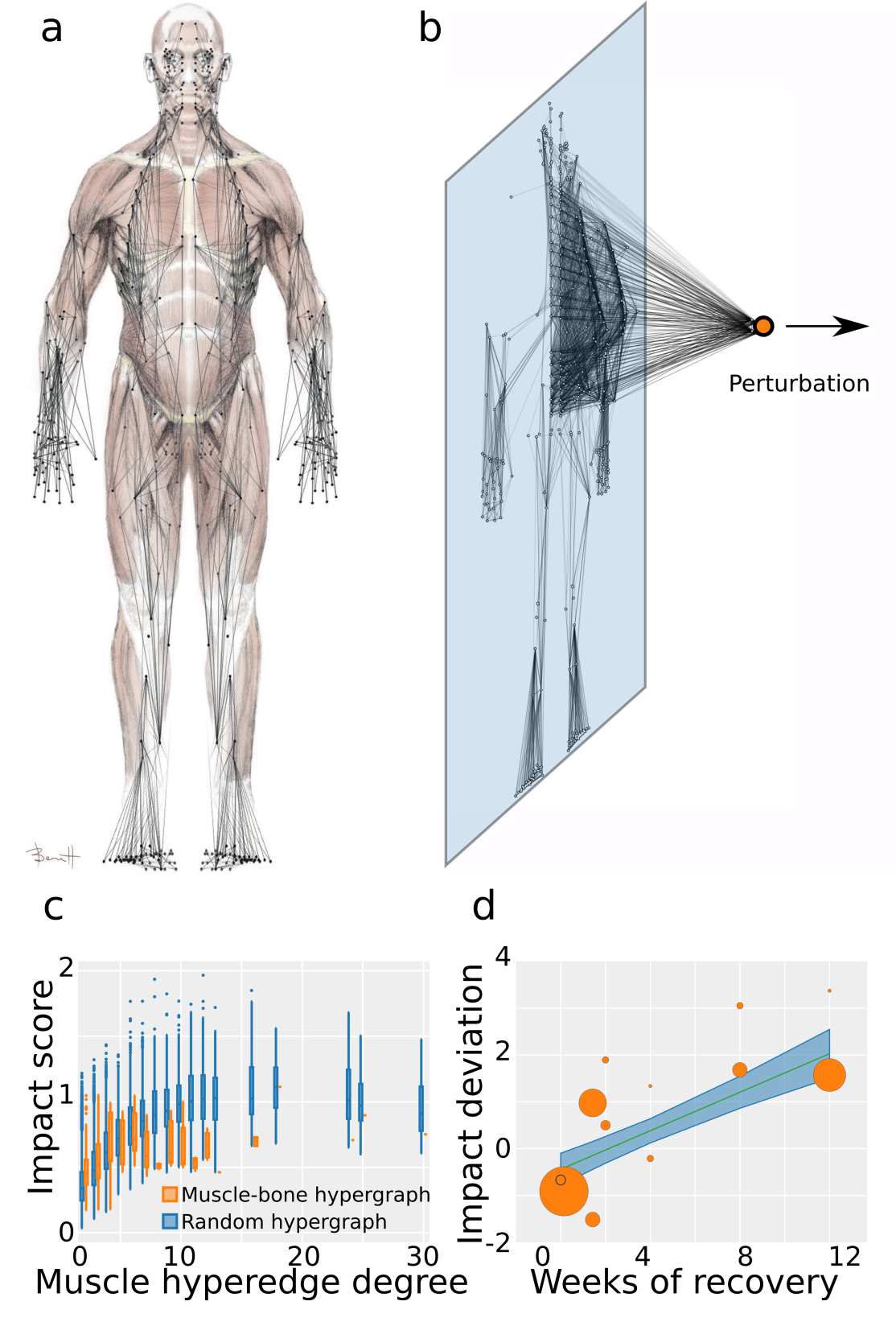}}
\caption{\linespread{1.5}\selectfont{} \textbf{Probing musculoskeletal function.}  (a) Visualization of the musculoskeletal network. (b) All nodes linked by a selected hyperedge are perturbed in a fourth spatial dimension as shown from a flattened image of the network. (c) The impact score plotted as a function of the hyperedge degree for a random hypergraph and the observed musculoskeletal hypergraph.  (d) Impact score deviation correlates with muscle recovery time following injury to muscles or muscle groups ($p = 6.29e-6$, $R^{2} = 0.854$). Data points are scaled according to number of muscles included.\label{fig2}}
\end{figure}

To probe the functional role of muscles within the musculoskeletal network (Fig.~\ref{fig2}a), we implemented a physical model in which bones form the core scaffolding of the body, while muscles fasten this structure together. Each node (bone) is represented as a mass whose spatial location and movement are physically constrained by the hyperedges (muscles) to which it is connected. Specifically, bones are points located at their center of mass derived from anatomy texts\cite{McMinn}, and muscles are springs (damped harmonic oscillators) connecting these points\cite{Zanin2008,Zanin2010}; for a hyperedge of degree $k$, we create $k(k-1)/2$ springs linking the $k$ nodes. That is, for a muscle connecting $k$ bones, we place springs such that each of the $k$ muscles has a direct spring connection to each of the other $k-1$ bones.

Next, we perturbed each of 270 muscles in the body and calculated their impact score on the network (see Methods and Fig.~\ref{fig2}b). As a muscle is physically displaced, it causes a rippling displacement of other muscles throughout the network. The impact score of a muscle is the mean displacement of all bones (and indirectly, muscles) resulting from its initial displacement. We observed a significant positive correlation between muscle degree and impact score (Pearson's correlation coefficient $r=0.45$, $p<0.00001$; Fig.~\ref{fig2}c), suggesting that hyperedge structure dictates the functional role of muscles in the musculoskeletal network. Muscles with a larger number of insertion and origin points have a greater impact on the musculoskeletal system when perturbed than muscles with few insertion and origin points\cite{Chopp2014}. 

To guide interpretation, it is critical to note that the impact score, while correlated with muscle degree, is not perfectly predicted by it (Fig.~\ref{fig2}c). Instead, the local network structure surrounding a muscle also plays an important role in its functional impact and ability to recover. To better quantify the effect of this local network structure, we asked whether muscles existed that had significantly higher or significantly lower impact scores than expected in a random network.  We defined a positive (negative) impact score deviation that measures the degree to which muscles are more (less) impactful than expected given a random network (see Methods). Interestingly, muscles with more impact than expected (positive impact score deviation) tend to be anatomically grouped in the arm and shoulder girdle. This localization is consistent with the fact that rotator cuff muscles are notoriously susceptible to injury, and that injuries to these muscles are particularly debilitating when they occur (Table~\ref{Table1}).

Is this mathematical model clinically relevant? Does the body respond differently to injuries to muscles with higher impact score than to muscles with lower impact score? To answer this question, we assessed the potential relationship between muscle impact and recovery time following injury. Specifically, we gathered data on athletic sports injuries and the time between initial injury and return to sport. Critically, we observed that recovery times were strongly correlated with impact score deviations of the individual muscle or muscle group injured ($p = 6.29\times10^{-6}$, $R^{2} = 0.854$; Fig.~\ref{fig2}d), suggesting that our mathematical model offers a useful clinical biomarker for the network's response to damage.

\subsection*{Control of the musculoskeletal network}

\begin{figure}
\centerline{\includegraphics[]{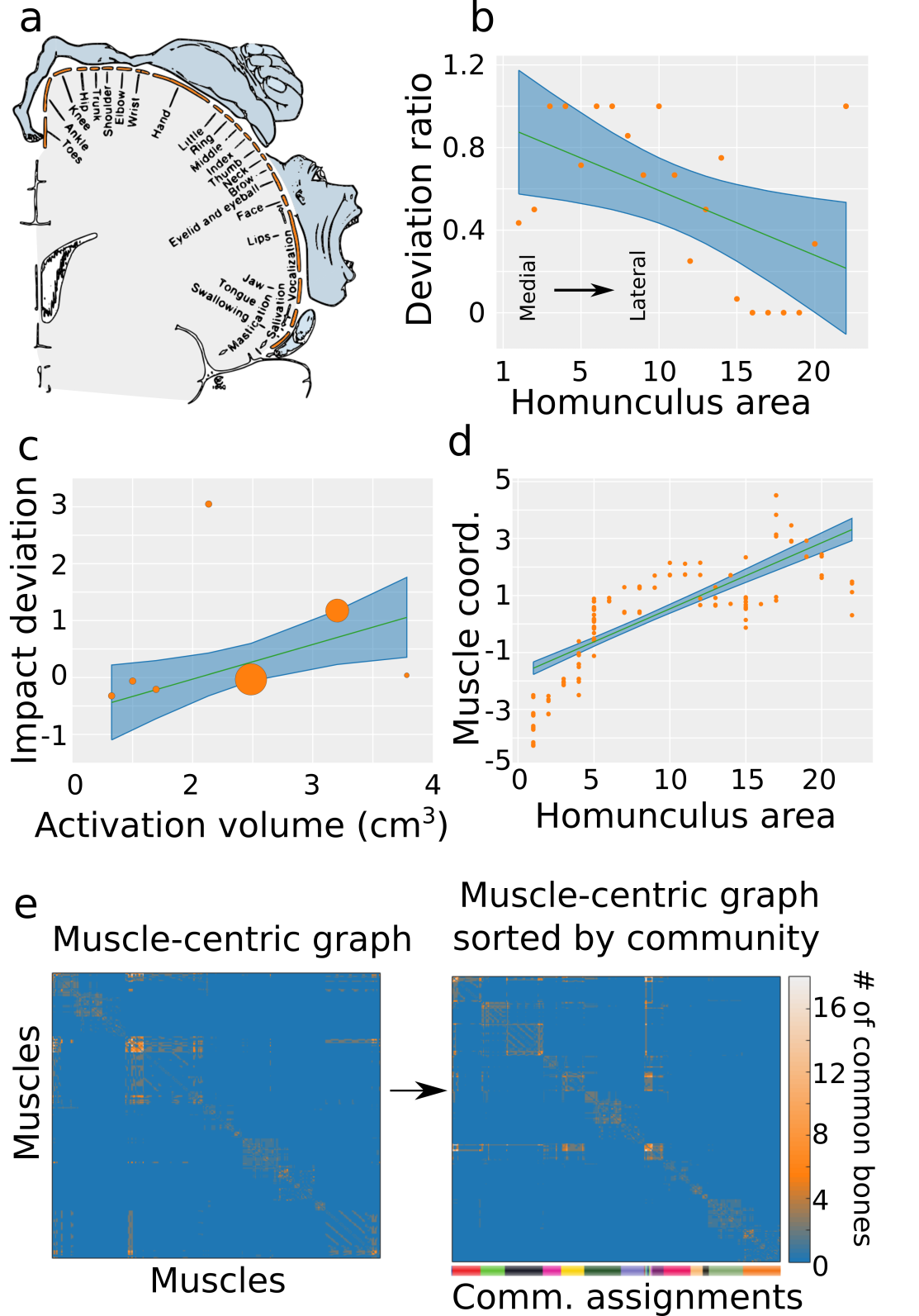}}
\caption{\linespread{1.5}\selectfont{} \textbf{Impact score and motor strip topology.}  (a) The motor cortex homunculus as first constructed by Penfield.  (b) Deviation ratio correlates significantly with Homuncular topology ($p = 0.0183$, $R^{2} = 0.26$), decreasing from medial (area 0) to lateral (area 22). (c) Impact score deviation significantly correlates with motor strip activation area ($p = 0.022$, $R^{2} = 0.728$), data points are sized according to the number of muscles required for the particular movement. (d) Correlation between the spatial ordering of Penfield's homunculus categories and the linear muscle coordinate ($p = 2.77e-47$, $R^{2} = 0.541$). (e) Community detection process.  The hypergraph is converted to a muscle-centric.  Modularity maximization extracts communities of densely interconnected muscles. The shaded areas indicate 95\% confidence intervals. \label{fig3}}
\end{figure}

What is the relationship between the functional impact of a muscle on the body and the neural architecture that affects control? Here, we interrogate the relationship between the musculoskeletal system and the motor cortex. We examine the cerebral cortical representation map area devoted to muscles with low \emph{versus} high impact by drawing on the anatomy of the motor strip represented in the motor homunculus\cite{Penfield} (Fig.~\ref{fig3}a), a coarse one-dimensional representation of the body in the brain\cite{Branco}. We observed that homunculus areas differentially control muscles with positive \emph{versus} negative impact deviation scores (Table~\ref{Table2}). Moreover, we found that homunculus areas controlling only positively (negatively) deviating muscles tend to be located medially (laterally) on the motor strip, suggesting the presence of a topological organization of a muscle's expected impact in neural tissue. To probe this pattern more deeply, for each homunculus area, we calculated a \emph{deviation ratio} as the percent of muscles that positively deviated from the expected impact score (i.e., a value of $1$ for brow, eye, face, and a value of $0$ for knee, hip, shoulder, see Table \ref{Table2}). We found that the deviation ratio was significantly correlated with the topological location on the motor strip ($p = 0.0183$, $R^{2} = 0.26$; Fig.~\ref{fig3}b).

As a stricter test of this relationship between a muscle's impact in the network and neural architecture, we collated data for the physical volumes of functional MRI-based activation on the motor strip that are devoted to individual movements (e.g., finger flexion or eye blinks). Activation volumes are defined as voxels that become activated (defined by blood-oxygen-level-dependent signal) during movement\cite{Indovina2000,Alkadhi2002}. Critically, we found that the functional activation volume independently predicts the impact score deviation of muscles (Fig.~\ref{fig3}c,  $p = 0.022$, $R^{2} = 0.728$), consistent with the intuition that the brain would devote more real estate in gray matter to the control of muscles that are more impactful than expected.

As a final test of this relationship, we asked whether the neural control strategy embodied by the motor strip is optimally mapped to muscle groups. We constructed a muscle-centric graph by connecting two muscles if they touch on the same bone (Fig.~\ref{fig3}e, right). We observed the presence of groups of muscles that were densely interconnected to one another, sharing common bones. We extracted these groups using a clustering technique designed for networks\cite{Porter2009,Fortunato2010}, which provides a data-driven partition of muscles into communities (Fig.~\ref{fig3}e, left). To compare this community structure present in the muscle network to the architecture of the neural control system, we considered each of the 22 categories in the motor homunculus\cite{Hosford} as a distinct neural community and compared these brain-based community assignments with the community partition calculated from the muscle network. Using the Rand coefficient\cite{Traud2010}, we found that the community assignments from both homunculus and muscle network were statistically similar ($z_{Rand}  > 10$), indicating a correspondence between the modular organization of the musculoskeletal system and structure of the homunculus. For example, the triceps brachii and the biceps brachii belong to the same homuncular category, and we found that they also belong to the same topological muscle network community. 

Next, because the homunculus has a linear topological organization, we asked whether the order of communities within the homunculus (Table~\ref{HomunculusTable}) was similar to a data-driven ordering of the muscle groups in the body, as determined by multidimensional scaling (MDS)\cite{Wang2011}. From the muscle-centric network (Fig.~\ref{fig3}e), we derive a distance matrix that encodes the smallest number of bones that must be traversed to travel from one muscle to another. An MDS of this distance matrix revealed a one-dimensional linear coordinate for each muscle, where topologically close muscles were close together and topologically distant muscles were far apart. We observed that each muscle's linear coordinate is significantly correlated with its homunculus category (Fig.~\ref{fig3}d, $p = 2.77 \times 10^{-47}$, $R^{2} = 0.541$), indicating an efficient mapping between the neural representation of the muscle system and the network topology of the muscle system in the body.

\section*{Discussion}

By representing the complex interconnectivity of the musculoskeletal system as a network of bones (represented by nodes) and muscles (represented by hyperedges), we gain valuable insight into the organization of the human body. The study of anatomical networks using similar methods is becoming more common in the fields of evolutionary and developmental biology\cite{Esteve2011}. However, the approach has generally been applied only to individual parts of the body -- including the arm\cite{Diogo2015}, the head \cite{Esteve2015}, and the spine\cite{Farfan1995} -- thereby offering insights into how that part of the organism evolved\cite{Rashevsky1954,Riedl1978}. Moreover, even when full body musculature \cite{Chiang2008} and the neuromusculoskeletal\cite{Murai2008} system more generally have been modeled, some quantitative claims can remain elusive in large part due to the lack of a mathematical language in which to discuss the complexity of the interconnection patterns. In this study, we offer an explicit and parsimonious representation of the \emph{complete} musculoskeletal system as a graph of nodes and edges, and this representation allows us to precisely characterize the network in its entirety.

When modeling a system as a network, it is important to begin the ensuing investigation by characterizing a few key architectural properties. One particularly fundamental measure of a network's structure is its degree distribution \cite{West2001}, which describes the heterogeneity of a node's connectivity to its neighbors in a manner that can provide insight into how the system formed \cite{albert2002statistical}. We observed that the degree distribution of the musculoskeletal system is significantly different from that expected in a random graph (Fig.~\ref{fig1}e), displaying fewer high degree nodes and an over abundance of low degree nodes. The discrepancy between real and null model graphs is consistent with the fact that the human musculoskeletal system develops in the context of physical and functional constraints that together drive its decidedly non-random architecture \cite{Glazier2012}. The degree distribution of this network displays a peak at approximately a degree of $2$, which is then followed by a relatively heavy tail of high degree nodes. The latter feature is commonly observed in many types of real-world networks \cite{Deng2011}, whose hubs may be costly to develop, maintain, and use\cite{Bullmore2012,Higurashi2008}, but play critical roles in system robustness, enabling swift responses\cite{Bullmore2012}, buffering environmental variation\cite{Levy2008}, and facilitating survival and reproduction\cite{He2006}. The former feature -- the distribution's peak -- is consistent with the intuition that most muscles within the musculoskeletal system connect with only two bones, primarily for the function of simple flexion or extension at a joint. By contrast, there are only a few muscles that require a high degree to support highly complex movements, such as maintaining the alignment and angle of the spinal column by managing the movement of many bones simultaneously.

The musculoskeletal network is characterized by a particularly interesting property that distinguishes it from several other real-world networks: the fact that it is embedded into 3-dimensional space \cite{barthelemy2011spatial}. This property is not observed in semantic networks \cite{gruenenfelder2016graph} or the world-wide web \cite{bornholdt2001world}, which encode relationships between words, concepts, or documents in some abstract (and very likely non-Euclidean) geometry. In contrast, the musculoskeletal system composes a volume, with nodes having specific coordinates and edges representing physically extended tissues. To better understand the physical nature of the musculoskeletal network, we examined the anatomical locations of muscles with varying degrees (Fig.~\ref{fig1}c). We observed that muscle hubs occur predominantly in the torso, providing dense structural interconnectivity that can stabilize the body's core and prevent injury \cite{Wilson2005}. Specifically, high degree muscles cluster about the body's midline, close to the spine, and around the pelvic and shoulder girdle, consistent with the notion that both agility and stability of these areas requires an ensemble of muscles with differing geometries and tissue properties \cite{Culham1993}. Indeed, muscles at these locations must support not only flexion and extension, but also abduction, adduction, and both internal and external rotation.

To better understand the functional role of a single muscle within the interconnected musculoskeletal system, we implemented a physics-based model of the network's impulse response properties by encoding the bones as point masses and the muscles as springs \cite{Blickhan1989}. While muscles of high degree also tended to have a large impact on the network's response (Fig.~\ref{fig2}c), there were several notable deviations from this tend (Table \ref{Table1}). The muscle with the most surprisingly low impact was located in the abdominal wall, where the transverse abdominal muscle, the rectus abdominis, and both the internal and external oblique muscles are tightly laminated together to perform distinct yet complementary functions \cite{DeTroyer1990}. This anatomical redundancy may explain the striking decrement in the impact score of the transverse abdominus in comparison to the null model: a muscle with less impact than expected may have several neighboring muscles in physical space that perform similar functional roles. Similarly, those muscles that had more impact than expected were largely located in the rotator cuff, an area known to have a highly sensitive muscle configuration based on clinical studies \cite{Yadav2009,Longo2011}.

While the network representation of a system can provide basic physical intuitions due to its parsimony and simplicity, it also remains agnostic to many details of the system's architecture and function. It is therefore a perennial question whether these first-principles models of complex systems can provide accurate predictions of real-world outcomes. We addressed this question by studying the relationship between the impact score of a muscle and the amount of time it takes for a person to recover from an injury. We quantified time of recovery by summing (i) the time to recover from the primary disability of the initial muscle injury, and (ii) the time to recover from any secondary disabilities resulting from altered usage of other muscles in the network due to the initial muscle injury \cite{Merrick1999}. We found that the deviation from expected impact score in a random network correlated significantly with time to recovery (Fig.~\ref{fig2}d), supporting the notion that focal injury can have extended impacts on the body due to the inherently interconnected nature of the musculoskeletal system. 

Indeed, muscular changes in one part of the body are known to effect other muscle groups. For example, strengthening hip muscles can lead to improved knee function following knee replacement \cite{Schache2016}. Alteration of muscular function in the ankle following sprains can cause altered hip muscle function \cite{Bullock1994,Bullock1994b}, and injury to limb muscles can lead to secondary injury of the diaphragm \cite{Road1998}. Our model offers a mathematically principled way in which to predict which muscles are more likely to have such a secondary impact on the larger musculoskeletal system, and which muscles are at risk for secondary injury given primary injury at a specific muscle site. It would be interesting in future to test whether these predictions could inform beneficial adjustments to clinical interventions by explicitly taking the risk of secondary injury to particular muscles into account. Previously, prevention of secondary muscle injury has been largely relegated to cryotherapy\cite{Kachanathu2013,Oliveira2006}, and has yet to be motivated by such a mechanistic model.

Given the complexity of the musculoskeletal network, and its critical role in human survival, it is natural to ask questions about how that network is controlled by the human brain. Indeed, the study of motor control has a long and illustrious history \cite{Franklin2011}, which has provided important insights into how the brain is able to successfully and precisely make voluntary movements despite challenges such as redundancies, noise \cite{Faisal2008}, delays in sensory feedback \cite{Matthews1991}, environmental uncertainty \cite{Scheidt2001}, neuromuscular nonlinearity \cite{Zajac1989}, and non-stationarity \cite{Dorfman1979}. Here we take a distinct yet complementary approach and ask how the topology of the musculoskeletal network maybe be mapped on to the topology of the motor strip within the cortex. We began by noting that the impact deviation of a muscle is positively correlated with the size of the cortical volume devoted to its control (Fig.~\ref{fig3}c). One interpretation of this relationship is that those muscles with greater impact than expected tend to control more complex movements, and therefore necessitate a larger number of neurons to manage those movements \cite{Catalan1998}. A second interpretation builds on an evolutionary argument that muscles with more impact need a greater redundancy in their control systems\cite{Guigon2007}, and this redundancy takes the form of more neurons.

Local cortical volumes aside \cite{Boonstra2015}, one might also wish to understand to what degree the larger-scale organization of the musculoskeletal network reflects the organization of the motor strip that controls it. Building on the recent application of community detection techniques to the study of skull anatomy \cite{Esteve2013,Esteve2014,Esteve2015}, we revealed the modular organization of the muscle network: groups of muscles in which the muscles in one group are more likely to connect to one another than to muscles in other groups. More intriguingly, we observed that muscle communities closely mimic the known muscle grouping of the motor strip (Fig.~\ref{fig3}e): muscles that tend to connect to the same bones as each other also tend to be controlled by the same portion of the motor strip. Furthermore, a natural linear ordering of muscle communities -- such that communities are placed close to one another on a line if they share network connections -- mimics the order of control in the motor strip (Fig.~\ref{fig3}d). These results extend important prior work suggesting that the two-dimensional organization of the motor strip is related to both the structural and functional organization of the musculoskeletal network \cite{Aflalo2006,Meier2008}. In fact, the results more specifically offer a network-level definition for optimal network control: the consistency of the linear map from musculoskeletal communities to motor strip communities. It would be interesting in future to test the degree to which this network-to-network map is altered in individuals with motor deficits or changes following stroke.

Finally, we interrogated the physical locations of the cortical control of impactful muscles. We observed that muscles with more impact than expected given a random graph tend to be controlled by medial points on the motor strip, while muscles with less impact than expected tend to be controlled by lateral points on the motor strip (Fig.~\ref{fig3}b). This spatial specificity indicates that the organization of the motor strip is constrained by both the physical layout of the body as well as aspects of how muscles function. Previous studies have examined a general temporal correspondence between cortical activity and muscle activity during movement \cite{Marsden2000}, but little is known about topological correspondence.

In summary, here we develop a novel network-based representation of the musculoskeletal system, construct a mathematical modeling framework to predict recovery, and validate that prediction with data acquired from athletic injuries. Moreover, we directly link the network structure of the musculoskeletal system to the organization of cortical architecture, suggesting an evolutionary pressure for optimal network control of the body. Our work directly motivates future studies to test whether faster recovery may be attained by not only focusing rehabilitation on the primary muscle injured, but also directing efforts towards muscles that the primary muscle impacts. Furthermore, our work supports the development of a predictive framework to determine the extent of musculoskeletal repercussions from insults to the motor cortex. An important step in the \emph{network science of clinical medicine}\cite{Barabasi2011}, our results inform the attenuation of secondary injury and the hastening of recovery.

\section*{Materials and Methods}

\subsection*{Network construction}

Using the traditional Hosford Muscle tables\cite{Hosford}, we construct a hypergraph by representing 173 bones (several of these are actually ligaments and tendons) as nodes and 270 muscles as hyperedges linking those nodes (muscle origin and insertion points are listed in Table \ref{BipartiteTable}). This hypergraph can also be interpreted as a bipartite network with muscles as one group and bones as the second group (Fig.~\ref{fig1}d). The 173 $\times$ 270 incidence matrix $\mathbf{C}$ of the musculoskeletal network is thus defined as $C_{ij} = 1$ if $v_i \in e_j$ and $0$ otherwise, where $V=\{v_1,\cdots,v_{173}\}$ is the set of nodes (bones) and $E=\{e_1,\cdots, e_{270}\}$ is the set of hyperedges (muscles). All analysis is applied to only one half (left or right) of the body because each cerebral hemisphere controls only the contralateral side of the body. In any figures where both halves of the body are shown, the second half is present purely for visual effect.

The bone-centric graph $\mathbf{A}$ and muscle-centric graph $\mathbf{B}$ (Fig. \ref{fig3}e) are simply the one-mode projections of $\mathbf{C}$. The projection onto bones is $\mathbf{A} = \mathbf{C^T}\mathbf{C}$, and the projection onto muscles is $\mathbf{B} = \mathbf{C}\mathbf{C^T}$. Then, the diagonal elements are set equal to zero, leaving us with a weighted adjacency matrix\cite{Newman2010}. We obtain estimated anatomical locations for the center of mass of each muscle (and bone) by examining anatomy texts\cite{McMinn} and estimating $x$, $y$, and $z$ coordinates for mapping to a graphical representation of a human body.

\subsection*{Network null models}

Random graphs are used in the current text as a null model against which to compare our real-world data. The random hypergraph is constructed first by randomly assigning edges such that each muscle has degree of two, in order to account for the fact that each muscle in the real graph has degree of at least 2. This results in a hypergraph composed of 540 connections. Because the true hypergraph has 1012 connections, 472 additional edges are uniformly randomly assigned within the hypergraph.

\subsection*{Calculation of impact score}
To measure the potential functional role of each muscle in the network, we use a classical perturbative approach. We perturb the location of a single muscle and observe the impact of this perturbation on the locations of all other muscles. Physically, to perturb a muscle, we displace all bones connected to that muscle by the same amount and in the same direction, and hold these bones fixed at their new location. In this way, we alter the location of a muscle. The system is then allowed to reach equilibrium. We fix bones at the midline and around the periphery in space in order to prevent the system from drifting. To quantify this impact, we define the movement of each node in the $i^{\mathrm{th}}$ hyperedge by the following equation:
\begin{equation} \label{eq:dampedharmonicoscillator}
I_{i} = q_i \sum_{j \neq i \in V} \left[\mathbf{A}_{ij} \hat{l}_{ij}(x_{ij}- \|\vec{l}_{ij}\| ) \right] - \beta\frac{d\vec{r}_i}{dt}= m_i\frac{d^2\vec{r}_i}{dt^2}, 
\end{equation}
where $l_{ij}$ is the displacement between nodes $i$ and $j$, $x_{ij}$ is the unperturbed distance between nodes $i$ and $j$, $m$ is the mass of the node (which, for simplicity and ease of interpretation, we have set equal to unity for all nodes in the network), $\beta=1$ is a damping coefficient, $r_i$ is the position of the $i^{th}$ node, $A$ is the weighted adjacency matrix of the bone-centric graph, and $q_i$ is the force constant for a spring in the $i^{\mathrm{th}}$ hyperedge. To normalize a node's restoring force, we let $q_i\propto{1/(k-1)}$. 

To measure the potential functional role of each muscle in the network, we perturb a muscle hyperedge and measure the impact of the perturbation on the rest of the network. Rather than perturbing the network in some arbitrary three-dimensional direction, we extend the scope of our simulation into a fourth dimension. When perturbing a muscle, we displace all of the nodes (bones) contained in that muscle hyperedge by a constant vector in the fourth dimension and hold them with this displacement (Fig.~\ref{fig2}b). The perturbation then ripples through the network of springs in response. We sequentially perturb each hyperedge, and define the impact score of this perturbation to be the mean distance moved by all nodes in the musculoskeletal network from their original positions. The displacement value is the summed displacement over all time points from perturbation onset to an appropriate cutoff for equilibration time.  Here, we solve for the equilibrium of the system by allowing dynamics to equalize over a sufficient period of time. We acknowledge that the equilibrium can be solved for using a steady-state, non-dynamic approach. We chose to use dynamics in this instance to more broadly support future applications

\subsection*{Impact score deviation}

For each muscle, we calculate an index that quantifies how much the impact score of that muscle deviates from expected given its hyperedge degree. To do this, we calculate the mean, standard deviation, and 95\% confidence intervals for each of the random hypergraph degree categories from a community of 100 random hypergraphs (Fig. \ref{fig2}c). The distance from a given muscle to the mean $\pm$ 95\% CI (whichever is closest) is calculated, and divided by the standard deviation of that random hypergraph degree distribution. In this way, we calculate a deviation from expected value, in standard deviations (similar to a $z$-score). Tables \ref{Table1} and \ref{Table2} contain the muscles which lie outside the 95\% confidence interval of deviation ratios relative to their hyperedge degree. For a given homunculus group, we calculate the \emph{deviation ratio} as the number of muscles with positive deviation divided by all muscles in the group.

\subsection*{Community detection}

Community detection is performed by maximizing the modularity quality function introduced by Newman\cite{Newman2006}. We maximize the following modularity quality function\cite{Newman2006}:
\begin{equation}
Q = \sum_{ij} B_{ij} - P_{ij} \delta(g_{i},g_{j}),
\end{equation}
where $P_{ij}$ is the expected weight of an edge in the Newman-Girvan null model, node $i$ is assigned to community $g_{i}$ and node $j$ is assigned to community $g_{j}$. By maximizing $Q$, we obtain a partition of nodes (muscles) into communities such that nodes within the same community are more densely interconnected than expected in a random network null model (Fig. \ref{fig3}e, left).

Here we also use a resolution parameter to control the size and number of communities detected. We utilize this mechanism in order to ensure that the number of communities detected matches the number of groups within the homunculus for straightforward comparison. We used a resolution parameter of $\gamma = 4.3$ to divide the muscle-centric matrix into 22 communities. In order to do this, we re-define the original muscle-centric matrix $A$ by following Jutla et. al.,\cite{Jutla2011}; specifically, we define $k = \sum_{i}A_{i,j}$, and then the adjusted matrix $B = A - \gamma\frac{k^Tk}{\sum_jk_j}$ is substituted to a locally greedy, Louvain-like modularity maximization algorithm\cite{Blondel2008}.

The above method of community detection is non-deterministic\cite{Good2010}. That is, the same solution will not be reached on each individual run of the algorithm. Therefore, one must ensure that the community assignments used are a good representation of the network, and not just a local minimum of the landscape. We therefore maximized the modularity quality function 100 times, obtaining 100 different community assignments. From this, a robust representative consensus community was found\cite{Bassett2013}. Figure \ref{figS1} illustrates how the detected communities change as a function of resolution parameter for the muscle-centric network.

\subsection*{Multidimensional scaling}

To conduct classical multidimensional scaling on the muscle-centric network, the weighted muscle-centric adjacency matrix was simplified to a binary matrix (all nonzero elements set equal to 1). From this, a distance matrix $\mathbf{D}$ was constructed whose elements $D_{ij}$ are equal to the length of the shortest path between muscles $i$ and $j$ and zero if no path exists. Classical multidimensional scaling is then applied to this distance matrix to yield its first principal component using the MATLAB function \emph{cmdscale.m}. To construct the binary matrix, a threshold of 0 was set, and all values above that were converted to 1. However, the relationship presented in Fig.~\ref{fig3}c is robust to this choice, and to verify this we explored a range of threshold values. The upper bound of the threshold range was established by determining the maximal value that would maintain a fully connected matrix, otherwise the distance matrix $\mathbf{D}$ would have entries of infinite weight. In our case, this value was $0.0556 \times Max(\mathbf{B})$. Within this range of thresholds (i.e. for all thresholds resulting in fully connected matrices), a significant fit can always be established. In addition to this, a method of constructing a distance matrix from a weighted adjacency matrix was employed in order to preclude thresholding (Fig. \ref{figS3}). Similarly, using this method yielded a significant linear relationship.

\subsection*{Muscle injury data}

We calculate the correlation between impact score and muscle injury recovery times. Injury recovery times were collected from the sports medicine literature and included injury to the triceps brachii and shoulder muscles\cite{Bateman1964}, thumb muscles\cite{Rettig2004}, latissimus dorsi and teres major\cite{Nagda2011}, biceps brachii\cite{Zafra2009}, ankle muscles\cite{McCollum2012}, neck muscles\cite{Torg1982}, jaw muscles\cite{Beachy2004}, hip muscles\cite{Niemuth2005}, eye/eyelid muscles\cite{Leivo2015}, muscles of the knee\cite{Ekstrand1982}, elbow\cite{Fleisig2012}, and wrist/hand\cite{Logan2004}. The recovery times and associated citations are listed in Table \ref{RecoveryTable}. If the literature reported a range of different severity levels and associated recovery times for a particular injury, the least severe level was selected. If the injury was reported for a group of muscles, rather than a single muscle, the impact score deviation for that group was averaged together. Data points for muscle groups were weighted according to the number of muscles in that group for the purpose of the linear fit. The fit was produced using the MATLAB function \emph{fitlm.m}, with option "Robust" set to "on."

\subsection*{Somatotopic representation area data}

We calculate the correlation between impact score deviation and the area of somatotopic representation devoted to a particular muscle group. The areas of representation were collected from two separate sources\cite{Indovina2000,Alkadhi2002}. The volumes and associated citations are listed in Table \ref{VolumeTable}. In both studies, subjects were asked to articulate a joint repetitively, and the volume of the areas of motor cortex that underwent the greatest change in BOLD signal were recorded. That volume was correlated with the mean impact of all muscles associated with that joint, as determined by the Hosford muscle tables. A significant linear correlation was found between the two measures by using the MATLAB function \emph{fitlm.m}, with option "Robust" set to "on."


\bibliography{bibfile_original}

\begin{thebibliography}{1}
\bibitem{use2e}
As the mark-up of the \TeX\ source for this document makes clear, your file
  should be coded in \LaTeX 2${\varepsilon}$, not \LaTeX\ 2.09 or an earlier
  release. Also, please use the \texttt{article} document class.
\bibitem{inclme}
Among whom are the author of this document. The ``real'' references and notes
  contained herein were compiled using B{\small{IB}}\TeX\ from the sample .bib
  file \texttt{scibib.bib}, the style package \texttt{scicite.sty}, and the
  bibliography style file \texttt{Science.bst}.
\end{thebibliography}

\bibliographystyle{Science}

\section*{Acknowledgments}
 We thank orthopedic surgeon John F. Perry, M.D., and professor of orthopedics Robert Mauck, M.D., for helpful comments on earlier versions of the manuscript. We also acknowledge the Penn Network Visualization Program. DSB and ACM would like to acknowledge support from the John D. and Catherine T. MacArthur Foundation, the Alfred P. Sloan Foundation, the Army Research Laboratory and the Army Research Office through contract numbers W911NF-10-2-0022 and W911NF-14-1-0679, the National Institute of Health (2-R01-DC-009209-11, 1R01HD086888-01, R01-MH107235, R01-MH107703, R01MH109520, 1R01NS099348 and R21-M MH-106799), the Office of Naval Research, and the National Science Foundation (BCS-1441502, CAREER PHY-1554488, BCS-1631550, and CNS-1626008). The content is solely the responsibility of the authors and does not necessarily represent the official views of any of the funding agencies.
 \\
The authors declare that they have no competing financial interests.

\beginsupplement
\newpage
\section*{Supplementary Materials}

\subsection*{Alternative perturbative approach.} In order to establish a measure of impact per muscle hyperedge, objects were displaced into a fourth spatial dimension to avoid making arbitrary choices within three dimensions. An alternative approach would be to perturb each muscle in each of three orthogonal directions, calculating impact each time, and calculating the vector sum of these three results. To answer the question of how these two approaches compare, we performed this experiment on the muscle-bone bipartite matrix to create two $270x1$ vectors, one encoding the impact scores via displacement in the fourth dimension, and one encoding the vector sum of the three orthogonal displacements. The two vectors were significantly correlated with each other (Pearson's $R = 0.9760$, $p = 1.6e-79$). 

\subsection*{Alternative random network null model.} To test robustness of the results to the choice of null model, we studied a second random network null model that maintained the hyperedge degree distribution of the real network. We constructed this model by randomly re-wiring each hyperedge in the graph, such that each hyperedge would maintain the same degree, but otherwise have connections assigned uniformly at random. Results from this degree-preserving null model are presented in Tables \ref{Table3} and \ref{Table4}. Note that the results are remarkably similar to those in Tables \ref{Table1} and \ref{Table2}, suggesting that our results are robust to the choice of null model.

\subsection*{Effect of resolution parameter in community detection.} 
\begin{figure}[H]
\centerline{\includegraphics[width=\textwidth]{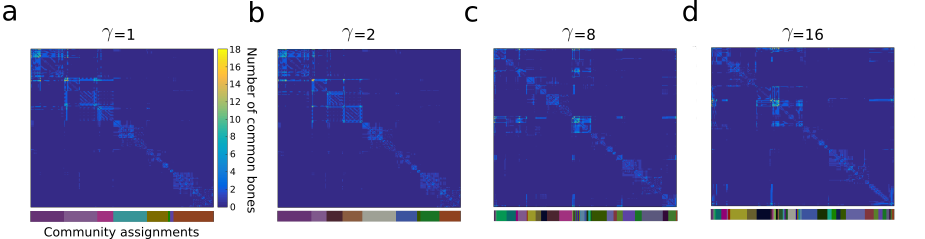}}
\caption{\linespread{1.5}\selectfont{} \textbf{Community detection with differing resolution parameters.}  This figure illustrates how the selection of the resolution parameter during community detection will change the number and size of communities detected. As the resolution parameter is increased, the size of individual communities decreases while the number of communities increases. (a-d) Community detection for the muscle-centric network, using $\gamma$ values of 1, 2, 8, and 16, respectively. The final community for each $\gamma$ is a consensus partition of 100 individual runs of the community detection algorithm.\label{figS1}}
\end{figure}

\begin{figure}[H]
\centerline{\includegraphics[width=.6\textwidth]{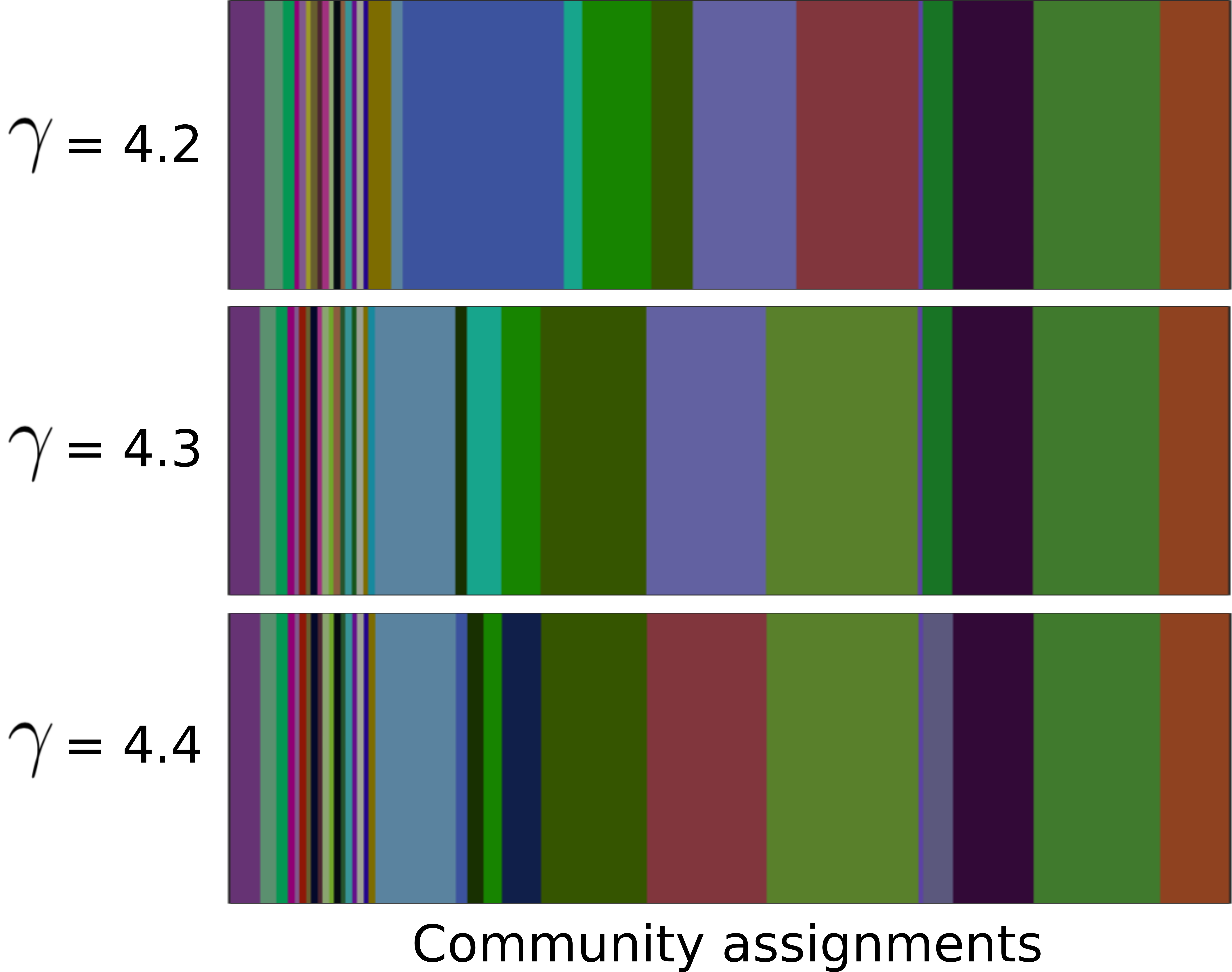}}
\caption{\linespread{1.5}\selectfont{} \textbf{Community detection with differing resolution parameters.}  This figure illustrates stability around the chosen tuning parameter of $\gamma = 4.3$. Here, we explore partitions generated from nearby resolution parameters $\gamma = 4.2$ and $\gamma = 4.4$. Visually, the three partitions appear to have similar structure. The two nearby partitions are also mathematically similar, with $z$-score of the Rand coefficient \cite{Traud2010} $\textup{zrand}(\gamma=4.2,\gamma=4.3)=105$, $\textup{zrand}(\gamma=4.3,\gamma=4.4)=110$, and $\textup{zrand}(\gamma=4.2,\gamma=4.4)=105$. The final community for each $\gamma$ is a consensus partition of 100 individual runs of the community detection algorithm. \label{figS2}}
\end{figure}

\begin{figure}[H]
\centerline{\includegraphics[]{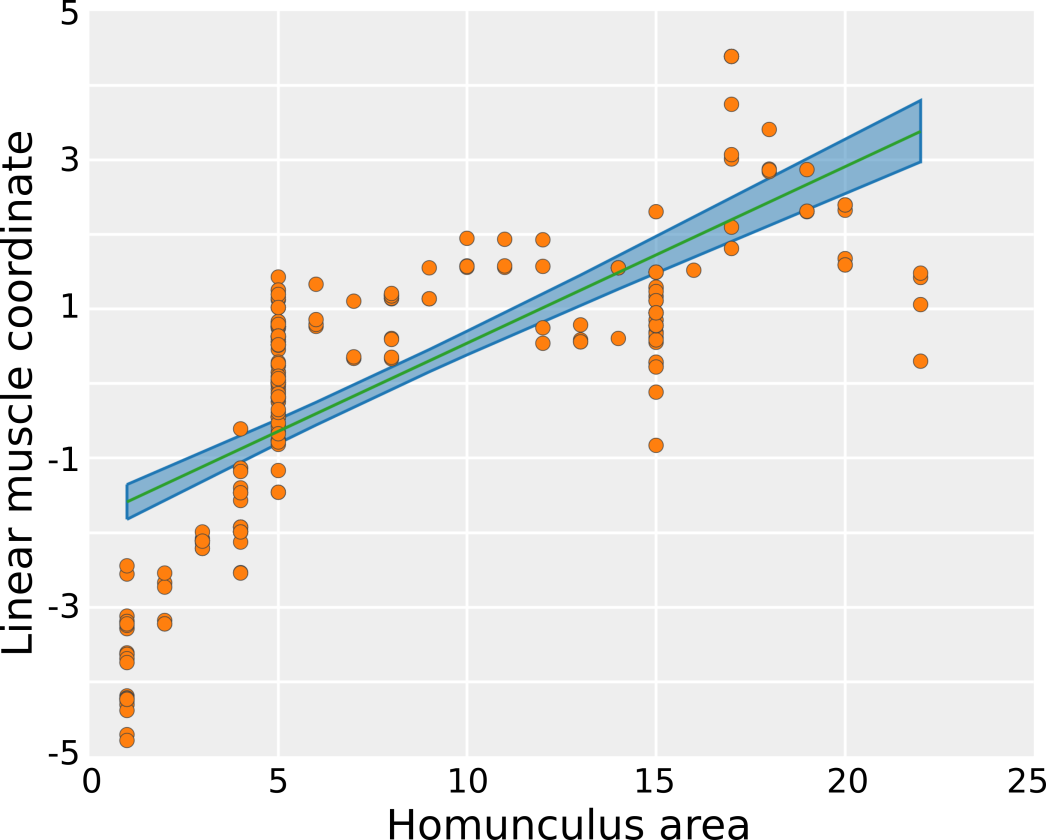}}
\caption{\linespread{1.5}\selectfont{} \textbf{Network topology and the homunculus.}  Linear muscle coordinates determined without thresholding using a weighted distance matrix (calculated using \emph{distance\_wei.m} included in the brain connectivity toolbox, https://sites.google.com/site/bctnet/). Without thresholding, a significant correlation also exists ($p = 6.21e-46$, $R^{2} = 0.531$).\label{figS3}}
\end{figure}

To detect network communities, we performed a modularity maximization approach as detailed in the Methods section. Importantly, the modularity quality function includes a resolution parameter, $\gamma$, that can be used to tune the relative size of the communities: smaller values of $\gamma$ identify larger communities and larger values of $\gamma$ identify smaller communities (Fig.~\ref{figS1}, Fig.~\ref{figS2}). We selected a single resolution parameter, $\gamma$ = 4.3, for the analysis presented in the main text to generate 22 communities, equal to the number of categories in the motor homunculus. To show that our results are robust to reasonable variations in this choice, we generated communities with nearby resolution parameters $\gamma$ =  4.2 and 4.4, which lead to the detection of 20 and 22 communities, respectively (Fig.~\ref{figS2}). These two partitions were statistically similar to the original partition, as tested by the $z$-score of the Rand coefficient \cite{Traud2010}: $z=105$ for $\gamma=4.2$ and $z=110$ for $\gamma=4.4$. 
\clearpage

\begin{table}[H]
\caption{\linespread{1.5}\selectfont{} \textbf{Muscles with greater and lesser impact than expected in a random hypergraph null model.} The muscles on the left side have less impact than expected given their hyperedge degree - their impacts are more than 1.96 standard deviations below the mean, indicating they lie outside the 95\% confidence interval of the distribution. The  muscles on the right side have more impact than expected given their hyperedge degree, their impacts are more than 1.96 standard deviations above the mean, ordered from most to least extreme. This table shows the muscles that had the greatest positive and greatest negative difference in impact relative to degree-matched controls.}
  \begin{tabular}{|c||c|c||c|c|}
    \hline
     Rank order &
      \multicolumn{2}{|c||}{Less impact than expected} &
      \multicolumn{2}{|c|}{More impact than expected}
       \\
       \hline
       \#&Hyperedge & Muscle name & Hyperedge & Muscle name\\
      \hline
    1&267&Transversus abdominus&20&Brachialis\\
    \hline
    2& & &22&Anconeus\\
    \hline
    3& & &18&Coracobrachialis\\
    \hline
    4& & &12&Teres minor\\
    \hline
    5& & &11&Infraspinatus\\
    \hline
    6& & &14&Subscapularis\\
    \hline
    7& & &13&Teres major\\
    \hline
    8& & &10&Supraspinatus\\
    \hline
     9& & &32&Extensor carpi radialus longus\\
    \hline
     10& & &161&Piriformis\\
    \hline
      11& & &31&Brachioradialis\\
    \hline
  \end{tabular}
  \label{Table1}
\end{table}

\begin{table}[H]
\caption{\linespread{1.5}\selectfont{} \textbf{Homunculus categories whose member muscles either all have more impact than expected or all have less impact than expected compared to random hypergraphs.} Categories on the left are composed entirely of muscles with less impact than expected compared to degree-matched controls. Categories on the right are composed entirely of muscles with more impact than expected compared to degree-matched controls.}
  \begin{tabular}{|c||c|c||c|c|}
    \hline
     Rank order &
      \multicolumn{2}{|c||}{Less impact than expected} &
      \multicolumn{2}{|c|}{More impact than expected}
       \\
       \hline
       \#&Homunculus category & Category name & Homunculus category & Category name\\
      \hline
    1&16&Brow&3&Knee\\
    \hline
    2&17&Eye muscles&4&Hip\\
    \hline
    3&18&Face muscles&6&Shoulder\\
    \hline
    4&19&Lip muscles&7&Elbow\\
    \hline
    5& & &10&Little finger\\
    \hline
    6& & &22&Swallowing\\
    \hline
      \end{tabular}
      \label{Table2}
\end{table}

\begin{table}[H]
\caption{Homunculus categories and their associated identification numbers.}
  \begin{tabular}{|l|c|}
    \hline
     Category ID & Category name\\
     \hline
     1 & Toes \\
     2 & Ankle \\
     3 & Knee \\
     4 & Hip \\
     5 & Trunk \\
     6 & Shoulder \\
     7 & Elbow \\
     8 & Wrist \\
     9 & Hand \\
     10 & Little finger \\
     11 & Ring finger \\
     12 & Middle finger \\
     13 & Index finger \\
     14 & Thumb \\
     15 & Neck \\
     16 & Brow \\ 
     17 & Eyelid and Eyeball \\
     18 & Face \\
     19 & Lips \\
     20 & Jaw \\
     21 & Tongue \\
     22 & Swallowing \\
     \hline
      \end{tabular}
      \label{HomunculusTable}
\end{table}

\begin{table}[H]
\caption{\linespread{1.5}\selectfont{} \textbf{Muscles with greater and lesser impact than expected in a randomly rewired hypergraphs.} The muscles on the left side have less impact than expected given their hyperedge degree - their impacts are more than 1.96 standard deviations below the mean, indicating they lie outside the 95\% confidence interval of the distribution. The  muscles on the right side have more impact than expected given their hyperedge degree, their impacts are more than 1.96 standard deviations above the mean, ordered from most to least extreme. This table shows the muscles that had the greatest positive and greatest negative difference in impact relative to degree-matched controls.}
  \begin{tabular}{|c||c|c||c|c|}
    \hline
     Rank order &
      \multicolumn{2}{|c||}{Less impact than expected} &
      \multicolumn{2}{|c|}{More impact than expected}
       \\
       \hline
       \#&Hyperedge & Muscle name & Hyperedge & Muscle name\\
      \hline
    1&72&Semispinalis thoracis&20&brachialis\\
    \hline
    2&61&Splenus capitis&22&Anconeus\\
    \hline
    3& & &18&Coracobrachialis\\
    \hline
    4& & &12&Teres minor\\
    \hline
    5& & &11&Infraspinatus\\
    \hline
    6& & &14&Subscapularis\\
    \hline
    7& & &13&Teres major\\
    \hline
    8& & &10&Supraspinatus\\
    \hline
     9& & &32&Extensor carpi radialus longus\\
    \hline
     10& & &161&Prirformis\\
    \hline
  \end{tabular}
  \label{Table3}
\end{table}

\begin{table}[H]
\caption{\linespread{1.5}\selectfont{} \textbf{Homunculus categories whose member muscles either all have more impact than expected or all have less impact than expected compared to randomly rewired hypergraphs.} Categories on the left are composed entirely of muscles with less impact than expected compared to degree-matched controls. Categories on the right are composed entirely of muscles with more impact than expected compared to degree-matched controls.}
  \begin{tabular}{|c||c|c||c|c|}
    \hline
     Rank order &
      \multicolumn{2}{|c||}{Less impact than expected} &
      \multicolumn{2}{|c|}{More impact than expected}
       \\
       \hline
       \#&Homunculus category & Category name & Homunculus category & Category name\\
      \hline
    1&16&Brow&3&Knee\\
    \hline
    2&17&Eye muscles&4&Hip\\
    \hline
    3&18&Face muscles&6&Shoulder\\
    \hline
    4&19&Lip muscles&7&Elbow\\
    \hline
    5& & &10&Little finger\\
    \hline
    6& & &22&Swallowing\\
    \hline
      \end{tabular}
      \label{Table4}
\end{table}

\begin{table}[H]
\caption{Muscle injury recovery times.}
  \begin{tabular}{|l|l|l|}
    \hline
     Muscle & Weeks of recovery & Source\\
     \hline
     Triceps brachii & 4 & Bateman (1962) \\
     Thumb muscles & 4 & Rettig (2004) \\
     Latissimus Dorsi & 12 & Nagda (2011) \\
     Biceps brachii & 12 & Zafra (2009) \\
     Ankle & 2 & McCollum (2012) \\
     Neck & 0.14 & Torg (1982) \\
     Jaw & 0 & Beachy (2004) \\
     Shoulder & 2 & Bateman (1962) \\
     Teres major & 12 & Nagda (2011) \\
     Hip & 12 & Niemuth (2005) \\
     Eye/eyelid & 1.4 & Leivo (2015) \\
     Knee & 8 & Ekstrand (1982) \\
     Elbow & 8 & Fleisig (2012) \\
     Wrist/hand & 1.4 & Logan (2004) \\
        \hline
      \end{tabular}
      \label{RecoveryTable}
\end{table}

\begin{table}[H]
\caption{Primary motor cortex somatotopic representation area sizes.}
  \begin{tabular}{|l|l|l|}
    \hline
     Muscle & Area ($mm^3$) & Source\\
     \hline
     Thumb & 1390 & Indovina (2000) \\
     Index & 1000 & Indovina (2000) \\
     Middle & 650 & Indovina (2000) \\
     Hand & 5566 & Alkadhi (2002) \\
     Fingers & 2972 & Alkadhi (2002) \\
     Wrist & 4409 & Alkadhi (2002) \\
     Elbow & 2267 & Alkadhi (2002) \\
             \hline
      \end{tabular}
      \label{VolumeTable}
\end{table}

{\tiny

\begingroup
\setlength{\LTleft}{-1.6cml}
\setlength{\LTright}{\LTleft}

\begin{longtable}{|p{2cm}|p{3cm}p{2cm}p{1.2cm}p{3cm}p{2cm}p{1.5cm}p{1.7cm}|}

\caption{\linespread{1.5}\selectfont{}  \normalsize{The muscles used in the current experiment and their attachments to bone. Attachments to bone are listed individually, and specific notes are provided when necessary.}}
\label{BipartiteTable}
\\\hline
\setcounter{LTchunksize}{1}

\bfseries Muscle & \bfseries Origin & \bfseries Notes & \bfseries No.~of~origins & \bfseries Insertions & \bfseries Notes & \bfseries No.~of~insertions & \bfseries Hyperedge~Degree
\begin{filecontents*}{grade.csv}
Muscle,Origin,NotesA,numOfOriginsTotal,Insertions,NotesB,numOfInsertionsTotal,HyperedgeDegree
Trapezius,1. external occipital protuberance,,22,1. posterior and lateral 1/3 of the clavicle,,3,25
,2. medial 1/3 of the superior nuchal line,,,2. medial margin of the acromion,,,
,3. ligamentum nuchae (surrounding the cervical spinous processes),,,3. superior spine of the scapula,,,
,4. spinous processes of C1-T12 (19 bones),considered as 19 origins,,,,,
Latissimus dorsi,1. spinous processes of T7-L5 (11 processes),considered as 11 origins,17,1. floor of the bicipital groove to the humerus,,1,18
,2. upper 2-3 sacral segments,considered as 1 origin: the sacrum,,,,,
,3. iliac crest,,,,,,
,4. lower 3 or 4 ribs,considered as 3 origins,,,,,
,5. inferior angle of the scapula,considered as: scapula,,,,,
Serratus posterior superior,1. spinous processes and supraspinous ligaments of C7-T2,considered as 3 origins,3,1. posterior aspect of ribs 2-5,considered as 4 insertions,4,7
Serratus posterior inferior,1. spinous processes and supraspinous ligaments T11-L2,considered as 4 origins,4,1. posterior aspect of ribs 9-12,considered as 4 insertions,4,8
Levator scapulae,1. posterior tubercles of the transverse processes of the upper 3 or 4 cervical vertebrae,considered as 4 origins,4,1. superior angle of scapula at and above scapular spine,considered as 1 insertion,1,5
Rhomboid minor,1. spinous process of C7 and T1,considered as 2 origins,3,1. medial margin of the scapula at the root of the spine,,1,4
,2. lower part of the ligamentum nuchae,,,,,,
Rhomboid major,1. spinous processes of T2-T5,considered as 4 origins,4,1. medial scapula from the scapular to the inferior angle,,1,5
Serratus anterior,1. outer surfaces and superior borders of the upper 8 or 9 ribs,considered as 8 origins,8,1. costal aspect of medial margin of the scapula,,1,9
Deltoid,1. anterior portion: anterior border of the lateral 1/3 of the clavicle,,3,1. deltoid tuberosity on the lateral surface of the shaft of the humerus,,1,4
,2. middle portion: lateral border of the acromion process of the scapula,,,,,,
,3. posterior portion: scapular spine,,,,,,
Supraspinatus,1. supraspinous fossa of the scapula,(scapula),1,1. uppermost of three facets of the greater tubercle of humerus,,1,2
Infraspinatus,1. infraspinous fossa of the scapula,(scapula),1,1. middle facet of greater tubercle of humerus,,2,3
Teres minor,1. middle half of the scapula's lateral margin,,1,1. lowest of three facets of the greater tubercle of humerus,,1,2
Teres major,1. lower third of the posterior surface of the lateral margin of the scapula,,1,1. medial lip of the bicipital groove fo the humerus (just medial to the insertion of latissimus dorsi),,1,2
Subscapularis,1. subscapular fossa on the anterior surface of the scapula,,1,1. lesser tubercle of humerus,,1,2
Subclavius,1. first rib about the junction of bone and cartilage,,1,1. lower surface of clavicle,,1,2
Pectoralis major,1. medial 1/3 of clavicle,,9,1. lateral lip of bicipital groove to the crest of the greater tubercle,,1,10
,2. anterior aspect of the sternum,,,,,,
,3. upper 6 costal cartilages,considered as 6 origins,,,,,
,4. aponeurosis of the external oblique,,,,,,
Pectoralis minor,1. outer surface of ribs 3-5 (may be variable),considered as 3 origins,3,1. medial aspect of coracoid process of the scapula,,1,4
Coracobrachialis,1. coracoid process of the scapula,,1,1. medial shaft of the humerus at about its middle,,1,2
Biceps brachii,1. lower 1/2 of anterior humerus,,2,1. radial tuberosity,,2,4
,2. short head-tip of the coracoid process of the scapula,,,2. bicipital aponeurosis,,,
Brachicalis,1. lower 1/2 of anterior humerus,,1,1. ulnar tuberosity,,1,2
Triceps brachii,1. long head: infraglenoid tubercle of the scapula,,2,1. posterior surface of the olecranon process of the ulna,,2,4
,2. lateral head: upper half of the posterior surface of the shaft of the humerus and the upper part of the lateral intermuscular septum,,,2. deep fascia of the antebrachium,,,
Anconeus,1. posterior surface of the lateral epicondyle of humerus,,1,1. lateral surface of olecranon extending to the lateral part of ulnar body,,1,2
Pronator teres,1. humeral head: a) upper portion of medial epicondyle via the CFT(common flexor tendon) b) medial brachial intermuscular septum,considered as 1 origins (CFT),3,1. lateral aspect of radius at the middle of the shaft (pronator tuberosity),,1,4
,2. ulnar head: coronoid process of ulna,,,,,,
,3. antebrachial fascia,,,,,,
Flexor carpi radialis,1. medial epicondyle of the humerus via the CFT,,2,1. base of the 2nd and sometimes 3rd metacarpals,considered as 2 insertions,2,4
,2. antebrachial fascia,,,,,,
Palmaris longus,1. medial epicondyle via the CFT,,2,1. central portion of the flexor retinaculum,,2,4
,2. antebrachial fascia,,,2. superficial portion of the palmar aponeurosis,,,
Flexor carpi ulnaris,1. humeral head: medial epicondyle via the CFT,,2,1. pisiform and hamate bones (via the pisohamateligament),considered as 2 insertions,3,5
,2. ulnar head: a) medial aspect of olecranon b) proximal 3/5 of dorsal ulnar shaft c) antebrachial fascia,considered as 1 origin (ulna),,2. base of the 5th metacarpal (via the pisometacarpal ligament),,,
Flexor digitorum superficialis,1. humeral-ulnar head: a) medial epicondyle via the CFT b) medial boarder of base of coronoid process of ulna c) medial (ulnar) collateral ligament d) antebrachial fascia,considered as 2 origins (ulna and CFT),3,1. both sides of the base of each middle phalanx of the 4 fingers,,4,7
,2. radial head: oblique line of radius along its anterior surface,,,,,,
Flexor digitorum profundus,1. anterior and medial surface of proximal 3/4 ulna,,1,distal phalanx of medial 4 digits (through the FDS tunnel),,4,5
,2. adjacent interosseous membrane,distal phalanx of medial 4 digits (through the FDS tunnel),,,,,
Flexor pollicis longus,1. middle anterior surface of the radius,,2,palmar aspect of base of the distal phalanx of thumb (deep to flexor retinaculum),,1,3
,2. interosseous membrane,,,,,,
,3. (may also originate from lateral boarder of coronoid process,,,,,,
,4. or medial epicondyle),,,,,,
Pronator quadratus,1. distal 1/4 anteromedial surface of ulna,,1,distal 1/4 anterolateral surface of radius,,1,2
Brachioradialis,1. upper lateral supracondylar ridge of humerus (between the triceps and brachialis muscles),,1,1. superior aspect of styloid process of radius,,2,3
,,,,3. antebrachial fascia,,,
Extensor carpi radialis longus,1. lateral intermuscular septum of humerus,,1,1. base of 2nd metacarpal,,1,2
Extensor carpi radialis brevis,1. lateral epicondyle via the CET (common extensor tendon),,3,1. base of 3rd metacarpal,,1,4
,2. radial collateral ligament,,,,,,
,3. antebrachial fascia,,,,,,
Extensor digitorum,1. lateral epicondyle via the CET,,2,1. base of middle phalanx of each of the four fingers (central band),,8,10
,2. antebrachial fascia,,,2. base of distal phalanx of each of the four finger bands (2 lateral bands),considered as 4 insertions unsure,,
Extensor digiti minimi,1. lateral epicondyle via the CET,,2,1. base of middle phalanx of the 5th digit (central band),,2,4
,2. antebrachial fascia,,,2. base of distal phalanx of the 5th digit (2 lateral bands),,,
Extensor carpi ulnaris,1. 1st head: lateral epicondyle via the CET,,3,1. medial side of base of the 5th metacarpal,,1,4
,2. 2nd head: posterior body of ulna,,,,,,
,3. antebrachial fascia,,,,,,
Supinator,1. lateral epicondyle of humerus,,5,1. proximal portion of anterolateral surface of the radius,,1,6
,2. supinator crest of ulna,,,,,,
,3. radial collateral ligament,,,,,,
,4. annular ligament,,,,,,
,5. antebrachial fascia,,,,,,
Abductor pollicis longus,1. posterior surfaces of ulna and radius,considered as 2 origins,4,1. lateral aspect of base of 1st metacarpal,,1,5
,2. interosseous membrane,,,,,,
,3. antebrachial fascia,,,,,,
Extensor pollicis brevis,1. posterior surfaces of radius (below abductor pollicis longus),,3,1. base of proximal phalanx of thumb (often a slip inserts into extensor pollicis longus tendon),,1,4
,2. interosseous membrane,,,,,,
,3. antebrachial fascia,,,,,,
Extensor pollicis longus,1. posterior surface of ulna,,3,1. distal phalanx of thumb,,1,4
,2. interosseous membrane,,,,,,
,3. antebrachial fascia,,,,,,
Extensor indicis,1. posterior surface of ulna (distal to extensor policis longus),,3,1. base of middle and distal phalanx of the index finger,,2,5
,2. interosseous membrane,,,,,,
,3. antebrachial fascia,,,,,,
Abductor pollicis brevis,1. distal border of flexor retinaculum,,2,1. lateral aspect of base of proximal phalanx of the thumb,,1,3
,2 trapezium (may be variable),,,,,,
Flexor pollicis brevis,1. superficial head: a) distal border of flexor retinaculum b) trapezium,considered as 2 origins,2,1. base of proximal phalanx of thumb,,1,3
Opponens pollicis,1. distal border of flexor retinaculum,,2,1. lateral aspect of the 1st metacarpal,,1,3
,2. trapezium,,,,,,
Adductor pollicis,1. transverse head: 3rd metacarpal,,5,1. medial aspect of the base of proximal phalanx,,2,7
,2. oblique head: a) base of 1st 2nd and 3rd metacarpals b) floor of carpal tunnel (trapezoid and capitate),considered as 4 additional origins,,2. medial sesamoid at McP (pisiform),,,
Palmaris brevis,1. medial margin of palmar aponeurosis,,1,1. may insert on the pisiform,,1,2
Abductor digiti minimi,1. pisiform and tendon of flexor carpi ulnaris,considered as 1 origin,1,1. medial aspect of the base of proximal phalanx of the 5th digit,,2,3
Flexor digiti minimi brevis,1. distal border of flexor retinaculum,,2,1. medial aspect of the base of proximal phalanx,,1,3
,2. hook of the hamate,,,,,,
Opponens digiti minimi,1. distal border of flexor retinaculum,,2,1. medial aspect of the 5th metacarpal,,1,3
,2. hook of the hamate,,,,,,
Palmar interossei (three of these per hand),1. from the side of the metacarpal that faces the midline - to adduct them,,1,1. on the base of the proximal phalanx of the digit of origin (same side toward the midline),,2,3
,,,,2. extensor hood of the same digit(s),,,
Dorsal interossei (three per hand),1. between each metacarpal,considered on a per finger basis,2,1. directly distal to the origin on the base of the proximal phalanx closest to the midline (to abduct them),,2,4
,,,,2. extensor hood of the same digit(s),,,
Lumbricals,1. tendon of flexor digtitorum: 1 and 2 have a single head of origin (from radial aspect of tendon) 3 and 4 have two heads of origin (each head from an adjacent tendon),1 (flexor digitorum profundus),1,extensor hood of digits 2-5,,1,2
Splenius capitis,1. lower portion of ligamentum nuchae,,9,1. superior nuchal line,(occipital bone),2,11
,2. spinous processes of C3-T3(4),considered as 8 origins,,2. mastoid process of temporal bone,,,
Splenius cervicis,1. spinous process of T3-T6,considered as 4 origins,4,1. superior nuchal line,,2,6
,,,,2. mastoid process of temporal bone,,,
Iliocostalis lumborum,1. sacrum,,9,1.  lower border of angles of ribs (5)6-12,considered as 7 insertions,7,16
,2. iliac crest,,,,,,
,3. spinous processes of lower thoracic and most lumbar vertebrae,7,,,,,
Iliocostalis thoracis,1. upper border of ribs 6-12 (medial to I. lumborum's insertion.),,6,1. lower border of angles of ribs 1-6 (sometimes transverse process of C7),considered as 6 insertions,6,12
Iliocostalis cervicis,1. angles of ribs 1-6,considered as 6 origins,6,1. transverse processes of C4-C6,considered as 3 insertions,3,9
Longissimus thoracis,1. sacrum,,2,1. transverse processes of all thoracic vertebrae,12,28,30
,2. iliac crest,,,2. all ribs between tubercles and angles,12,,
,3. spinous processes of lower thoracic and most lumbar vertebrae,all considered insertions,,3. transverse processes of upper lumbar vertebrae,4,,
Longissimus cervicis,1. transverse processes of T1-T5(6),considered as 5 origins,5,1. transverse processes of C2-C6,considered as 5 insertions,5,10
Longissimus capitis,1. transverse and articular processes of middle and lower cervical vertebrae,considered 4 origins,8,1. posterior aspect of mastoid process of temporal bone,,1,9
,2. transverse processes of upper thoracic vertebrae,considered as 4 origins,,,,,
Spinalis thoracis,1. sacrum,,10,1. spinous processes T3(4)-T8(9),considered as 6 insertions,6,16
,2. iliac crest,,,,,,
,3. spinous processes of lower thoracic and most lumbar vertebrae,considered 8 origins,,,,,
Spinalis cervicis,1. spinous processes of C6-T2,considered as 4 origins,4,1. spinous processes of C2 (and possibly extend to C3 or C4),considered as 1 insertion,1,5
Spinalis capitis,1. spinous processes of lower cervical and upper thoracic vertebrae,considered 7 origins,7,1. between superior and inferior nuchal lines of occipital bone,,1,8
Semispinalis thoracis,1. transverse processes of T6-T12 vertebrae,considered as 7 origins,7,1. spinous processes of upper thoracic and lower cervical vertebrae,considered as 6 insertions,6,13
Semispinalis cervicis,1. transverse processes of T1-T6 vertebrae and can go down to lower thoracic,considered as 6 origins,6,1. spinous processes of C2-T5(6),considered as  6 new insertions,6,12
Semispinalis capitis,1. transverse processes of T1-T6,considered as 6 origins,10,1.  between superior and inferior nuchal lines of occipital bone,,1,11
,2. articular processes of C4-C7,considered as 4 origins,,,,,
Multifidus,1. cervical region: from articular processes of lower cervical vertebrae,considered as 6 origins,24,1. spinous process of all vertebrae extending from L5 - C2 (skipping 1-3 segments),0 new insertions,0,24
,2. thoracic region: from transverse processes of all thoracic vertebrae,considered as 12 origins,,,,,
,3. lumbar region: lower portion of dorsal sacrum and PSIS and deep surface of tendenous origin of erector spinae and mamillary processes of all lumbar vertebrae,considered as L1-L5 and sacrum,,,,,
Long rotators,1. transverse process of one vertebra,,1,1. skips one vertebra to insert on the base of spinous process of vertebra above,,1,2
Interspinalis,1. spinous processes of each vertebra,,1,1.  to the spinous process of vertebra immediately above,,1,2
Intertransversi,"1. cervical region: A) from the anterior tubercle of transverse process
             B) from the posterior tubercle of transverse process",(from A to A and B to B) considered as1 origins,1,1. cervical region: A) to the anterior tubercle immediately above B) to the posterior tubercle immediately above,(from A to A and B to B) considered as 1 insertions,1,2
,"3. lumbar region: A) lateral aspect of the transverse process
               B) mamillary process",(from A to A and B to B) considered as 1 origins,,3.lumber region: A) lateral aspect of the transverse process immediately above B) to the accessory process on the vertebra immediately above,(from A to A and B to B) considered as 2 insertions,,
Obliquus capitis inferior,1. spinous process of axis (C2),,1,1. transverse process of atlas (C1),,1,2
Obliquus capitis superior,1. transverse process of atlas (C1),,1,1. between superior and inferior nuchal line of occiput,,1,2
Rectus capitis posterior major,1.  spinous process of axis (C2),,1,1. inferior nuchal line (lateral to minor),,1,2
Rectus capitis posterior minor,1. posterior tubercle of atlas (C1),,1,1. inferior nuchal line (adjacent to midline),,1,2
Longus colli,1. lower anterior vertebral bodies and transverse processes,considered as 6 origins (C5-T3),6,1. anterior vertebral bodies and transverse processes several segments above,1 (Atlas),1,7
Longus capitis,1. upper anterior vertebral bodies and transverse processes,considered as 4 origins (C3-C6),4,1. anterior vertebral bodies and transverse processes several segments above,,1,5
Rectus capitis anterior,1. anterior base of the transverse process of the atlas,,1,1. occipital bone anterior to foramen magnum,,1,2
Rectus capitis lateralis,1. transverse process of the atlas,,1,1. jugular process of the occipital bone,,1,2
Anterior scalene,1. anterior tubercles of transverse processes of C3-C6,considered as 4 origins,4,1. 1st rib,,1,5
Scalenus minimus (may be absent),1. anterior tubercles of transverse processes of C6 and 7,considered as 2 origins,2,1.1st rib and/or supraplural membrane,considered as 1 insertion,1,3
Middle scalene,1. transverse processes of all cervical vertebrae,considered as 7 origins,7,1. 1st rib,,1,8
Posterior scalene,1. posterior tubercles of transverse processes of C5 and C6,considered as 2 origins,2,1. 2nd and/or 3rd rib,considered as 1 insertion,1,3
Sternocleidomastoid,1. manubrium of sternum,,2,1. mastoid process of temporal bone,,1,3
,2. medial portion of clavicle,,,,,,
Platysma,1. subcutaneous skin over delto-pectoral region,,1,1. invests in the skin widely over the mandible,1,1,2
Sternohyoid,1. posterior aspect of manubrium,,2,1. body of hyoid,,1,3
,2. sternal end of clavicle,,,,,,
Omohyoid,1. superior belly: hyoid bone (lateral to sternohyoid),,2,1. both bellies meet at the clavicle and are held to the clavicle by a pulley tendon,,1,3
,2. inferior belly: superior scapular border (medial to suprascapular notch),,,,,,
Sternothyroid,1. posterior aspect of manubrium,,1,1. oblique line of thyroid cartilage,,1,2
Thyrohyoid,1. oblique line of thyroid cartilage,,1,1. body of hyoid,,1,2
Stylohyoid,1. styloid process of temporal bone,,1,1. lateral margin of hyoid (near greater horn),,1,2
Digastric,1. post belly: mastoid process of temporal bone,,2,1. both bellies meet and attach at the lateral aspect of body of hyoid  by a pulley tendon,,1,3
,2. anterior belly: digastric fossa of internal mandible,,,,,,
Mylohyoid,1. inner surface of mandible off the mylohyoid line,,1,1. body of hyoid,,1,2
Geniohyoid,1. inner surface of the mandible off the mental spines,,1,1. body of hyoid (paired muscles separated by a septum),,1,2
Occipitalis (2 bellies),1. lateral 2/3 of superior nuchal line,,2,1.  galea aponeurosis over the occipital bone,,1,3
,2. external occipital protuberance,,,,,,
Frontalis (2 bellies),1. galea aponeurosis anterior to the vertex,,1,1. skin above the nose and eyes,,1,2
Orbicularis oculi,1. orbital portion: nasal process of frontal bone,,3,1. circumferentially around orbit meeting in palpebral raphe,,1,4
,2. palpebral portion: palpebral ligament,,,,,,
,3. lacrimal portion: lacrimal crest of lacrimal bone,,,,,,
Corrugator supercilii,1. frontal bone just above the nose,,1,1. skin of the medial portion of the eyebrows,,1,2
Orbicularis oris,1. alveolar border of maxilla,,2,1. circumferentially around mouth,,1,3
,2. lateral to midline of mandible,,,2. blends with other muscles,,,
Levator labii superioris alaeque nasi,1. frontal process of maxilla,,1,1. upper lip muscles,,2,3
,,,,2. nasal cartilage,,,
Levator labii superioris,1. medial 1/2 of infraorbital margin,,1,1. upper lip muscles,,1,2
Zygomaticus minor,1. zygomatic bone posterior to maxillary-zygomatic suture,,1,1. skin of the upper lip,,1,2
Zygomaticus major,1. anterior to zygomatic-temporal suture,,1,1. modiolus (angle of the mouth),,1,2
Levator anguli oris,1. maxilla inferior to infraorbital foramen,,1,1. modiolus (angle of the mouth),,1,2
Buccinator,1. posterior alveolar process of maxilla,,3,1. modiolus,,1,4
,2. posterior alveolar process of mandible,,,,,,
,3. along the pterygomandibular raphe,,,,,,
Depressor anguli oris,1. along the oblique line of mandible,,1,1. modiolus,,1,2
Masseter,1. Superficial: a) zygomatic process of the maxilla b) inferior border of zygomatic arch,,2,1. Superficial: a) angle of mandible b) lateral surface of mandibular ramus,considered as 1 insertion (mandible),1,3
,2. Intermediate: inner surface of zygomatic arch,,,,,,
Medial pterygoid,1. medial surface of lateral pterygoid plate of the sphenoid,,3,1. inner surface of mandibular ramus,,1,4
,2. palatine bone,,,,,,
,3. pterygoid fossa,,,,,,
Lateral pterygoid,1. Superior head: lateral surface of the greater wing of the sphenoid,,2,1. neck of the mandibular condyle,,2,4
Levator palpebrae superioris,1. inferior aspect of the lesser wing of sphenoid (adjacent to the common annular tendon),,1,1. medial and lateral walls of the orbit,,2,3
,,,,2. superior tarsus,,,
Lateral rectus,1. common annular tendon (which comes off the body and lesser wing of sphenoid),,2,1. posterior to the sclerocorneal junction (each muscle inserting along its own directional axis),,1,3
,2. margins of the optic canal,,,,,,
Medial rectus,1. common annular tendon (which comes off the body and lesser wing of sphenoid),,2,1. posterior to the sclerocorneal junction (each muscle inserting along its own directional axis),,1,3
,2. margins of the optic canal,,,,,,
Superior rectus,1. common annular tendon (which comes off the body and lesser wing of sphenoid),,2,1. posterior to the sclerocorneal junction (each muscle inserting along its own directional axis),,1,3
,2. margins of the optic canal,,,,,,
Inferior rectus,1. common annular tendon (which comes off the body and lesser wing of sphenoid),,2,1. posterior to the sclerocorneal junction (each muscle inserting along its own directional axis),,1,3
,2. margins of the optic canal,,,,,,
Superior oblique,1. body of sphenoid,,1,1.  upper lateral quadrant of the posterior half of the sclera (via the trochlea as a pulley),,1,2
Inferior oblique,1. orbital surface of maxilla (Annulus of Zinn),,1,1. lower lateral quadrant of the posterior half of the sclera (via the suspensory ligament as a pulley),,1,2
Tensor fascia lata,1. anterior aspect of iliac crest,,2,1. anterior aspect of IT band below greater trochanter,,1,3
Gluteus maximus,1. outer rim of ilium (medial aspect),,4,1. IT band (primary insertion),,2,6
,2. dorsal surface of sacrum and coccyx,2 origins,,2. gluteal tuberosity of femur,,,
,3.  sacrotuberous ligament,,,,,,
Gluteus medius,1. outer aspect of ilium (between iliac crest and anterior and posterior gluteal lines),,2,1. superior aspect of greater trochanter,,1,3
,2. upper fascia (AKA gluteal aponeurosis),,,,,,
Gluteus minimus,1. outer aspect of ilium (between anterior and inferior gluteal lines),,1,1. greater trochanter (anterior to medius) of femur,,1,2
Piriformis,1. pelvic surface of sacrum (anterior portion),,1,1. medial surface of greater trochanter (through greater sciatic foramen),,1,2
Superior gemellus,1. ischial spine,,1,1. medial aspect of greater trochanter via upper tendon of obturator internus,,1,2
Obturator internus,1. internal aspect margins of obturator foramen,,2,1. medial aspect of greater trochanter (through lesser sciatic foramen),,1,3
,2. obturator membrane,,,,,,
Inferior gemellus,1. ischial tuberosity,,1,1. medial aspect of greater trochanter via lower tendon of obturator internus,,1,2
Quadratus femoris,1. lateral aspect of ischial tuberosity,,1,1. quadrate line (along posterior aspect of femur and intertrochanteric crest),,1,2
Semitendinosus,1. ischial tuberosity,,1,1. medial aspect of tibial shaft,,2,3
,,,,2. contributes to the pez anserine,,,
Semimembranosus,1. ischial tuberosity,,1,1. posterior medial aspect of medial tibial condyle,,1,2
Biceps femoris,1. long head: ischial tuberosity,,2,1. head of fibula,,2,4
,2. short head: lateral lip of linea aspera and the lateral intermuscular septum (femur),,,2. maybe to the lateral tibial condyle,,,
Adductor longus,1. anterior surface of pubis just inferior to the pubic tubercle,,1,1. medial lip of linea aspera on middle half of femur,,1,2
Adductor brevis,1. body and inferior ramus of pubis,,1,1. superior portion of linea aspera,,1,2
Adductor magnus,1. anterior fibers: inferior pubic ramus,,2,1. proximal 1/3 of linea aspera,femur,2,4
,2. oblique fibers: ischial ramus,,,,,,
Gracilis,1. body of pubis and inferior pubic ramus,,1,1. medial surface of proximal tibia inferior to tibial condyle,,2,3
,,,,2. contributes to the pez anserine,,,
Obturator externus,1. external surface of obturator membrane,,1,1. trochanteric fossa of femur,,1,2
,,,,,,,
Sartorius,1. anterior superior iliac spine (ASIS),illium,1,1. upper medial surface of body of tibia,,2,3
,,,,2. contributes to pez anserine,,,
Rectus femoris,1. anterior head: anterior inferior iliac spine (AIIS),illium,1,1. common quadriceps tendon into patella,,2,3
,,,,2. tibial tuberosity via patellar ligament,,,
Vastus lateralis,1. greater trochanter,,1,1. common quadriceps tendon into patella,,2,3
,,,,2. tibial tuberosity via patellar ligament,,,
Vastus intermedius,1. anterior lateral aspect of the femoral shaft,,1,1. common quadriceps tendon into patella,,2,3
,,,,2. tibial tuberosity via patellar ligament,,,
Vastus medialis,1. intertrochanteric line of femur,,1,1. common quadriceps tendon into patella,,2,3
,,,,2. tibial tuberosity via patellar ligament,,,
Articularis genus,1 .distal portion of anterior femoral surface close to the knee,,1,1. synovial membrane of the knee joint,,1,2
Psoas major,1. transverse processes of L1-L5,considered as 5 origins,6,1. iliopsoas tendon to the lesser trochanter of the femur,treated as femur,1,7
,2. vertebral bodies of T12-L4 and the intervening intervertebral discs,considered as 1 new origin,,,,,
Iliacus,1. inner surface of upper iliac fossa,,1,1. iliopsoas tendon to the lesser trochanter of the femur,,1,2
Pectineus,1. pectineal line of the pubis,,1,1. the pectineal line of the femur,,1,2
Gastrocnemius,1. medial head: just above medial condyle of femur,,1,1. calcaneus via lateral portion of calcaneal tendon,,1,2
Soleus,1. upper fibula,,2,1. calcaneus via medial portion of calcaneal tendon,,1,3
,2. soleal line of tibia,,,,,,
Plantaris,1 .above the lateral head of gastrocnemius on femur,,1,1. calcaneus medial to calcaneal tendon or blending with the calcaneal tendon,,1,2
Popliteus,1. lateral femoral condyle,,2,1. posterior tibial surface above the soleal line,,1,3
,2. lateral meniscus,,,,,,
Flexor digitorum longus,1. posterior surface of tibia,,2,1. plantar surface of bases of the 2-5th distal phalanges,,4,6
,2. crural fascia,,,,,,
Tibialis posterior,1. posterior and proximal tibia,,3,1. navicular tuberosity (principle),1,9,12
,2. interosseous membrane,,,2. all 3 cuneiforms (plantar surface),3,,
,3. medial surface of fibula,,,3. bases of 2nd-4th metatarsals,3,,
,,,,4. cuboid,1,,
,,,,5. sustentaculum tali of calcaneus,1,,
Flexor hallucis longus,1. posterior and inferior 2/3 of fibula,,3,1. plantar surface of distal phalanx of hallux,,1,4
,2. interosseous membrane,,,,,,
,3. crural fascia and posterior intermuscular septum,,,,,,
Peroneus longus,1. head of the fibula,,1,1. plantar surface of cuboid,,5,6
,,,,2. base of 1st and (2nd) metatarsal,just 1st included,,
,,,,3. plantar surface of medial cuneiform,3,,
Peroneus brevis,1. distal 2/3 of lateral fibula,,1,1. tuberosity on lateral aspect of base of 5th metatarsal,,1,2
Tibialis anterior,1. lateral tibial condyle,,3,1. medial and plantar surface of base of 1st metatarsal,,4,7
,2. interosseous membrane,,,2. medial and plantar surface of the cuneiform,3,,
,3. anterior intermuscular septum and crural fascia,,,,,,
,,,,,,,
Extensor hallucis longus,1. medial aspect of the fibula,,3,1. dorsal surface of base of proximal and distal phalanx of hallux,2,2,5
,2. interosseous membrane,,,,,,
,3. crural fascia,,,,,,
Extensor digitorum longus,1. lateral condyle of the tibia,,4,1. dorsal surface of the bases of the middle and distal phalanxes of the 2nd-5th rays (via 4 tendons and giving a fibrous expansion),8,8,12
,2. upper anterior surface of fibula,,,,,,
,3. interosseous membrane,,,,,,
,4. crural fascia,,,,,,
Peroneus tertius,1. distal 1/3 of anterior fibula,,1,1. dorsal surface of base of 5th metatarsal,,1,2
Abductor hallucis,1. medial process of calcaneal tuberosity,,3,1. medial aspect of base of proximal phalanx of hallux,,1,4
,2. flexor retinaculum,,,,,,
,3. plantar aponeurosis,,,,,,
Flexor digitorum brevis,1. medial process of calcaneal tuberosity,,2,1. both sides of the bases of the middle phalanx of rays 2-5 (each of the 4 tendons splits forming tunnel for FDL),,4,6
,2. plantar aponeurosis,,,,,,
,,,,,,,
Abductor digiti minimi,1. lateral and medial processes of the calcaneal tuberosity,,2,1. lateral aspect of base of proximal phalanx of 5th ray,,1,3
,2. plantar aponeurosis,,,,,,
Abductor ossis metatarsi quinti,1. from fibers of abductor digiti minimi,,1,1. into the 5th metatarsal,,1,2
Quadratus plantae,1. medial head: medial calcaneus,,1,1. lateral margin of tendon of flexor digitorum longus (FDL) (may send slips into the distal tendons),,1,2
Lumbricals,1. from 1st tendon of FDL: medial aspect of tendon to 2nd ray,(FDL and proximal phalanx),2,1. extensor tendons of EDL on dorsal foot,,1,3
Flexor hallucis brevis,1. medial aspect of the cuboid,,2,1. medial aspect of base of proximal phalanx of hallux,,1,3
,2. lateral cuneiform (3rd),,,,,,
Adductor hallucis,1. oblique head: base of 2nd-4th metatarsals and long plantar ligament,3,3,1. lateral aspect of base of proximal phalanx of hallux,,1,4
Flexor digiti minimi brevis,1. base of 5th metatarsal,,1,1. lateral aspect of base of proximal phalanx of 5th ray,,,
Plantar interossei (3 muscles),1. medial aspect of 3rd-5th metatarsals (each muscle has a single head),,1,1. medial aspect of base of proximal phalanx of the same ray (of 3rd-5th rays),,1,2
Dorsal interossei (4 muscles),1.  from both metatarsals between which they lie,,2,1. base of proximal phalanx closest to the axis of the foot (2nd ray),,1,3
Extensor hallucis brevis,1. upper anterolateral calcaneus,,1,1. base of proximal phalanx of hallux,,1,2
Extensor digitorum brevis,1. upper anterolateral calcaneus,,1,1. middle and distal phalanges of 2nd-4th rays (via EDL),considered as 6 insertions,6,7
Cricothryoid,1. arch of cricoid cartilage,,1,1. lower border and lower medial surface of thyroid cartilage,,1,2
Posterior cricoarytenoid,1. posterior surface of cricoid cartilage at midline,,1,1. muscular process of the arytenoid cartilage,,1,2
Lateral cricoarytenoid,1. upper border of lateral aspect of the cricoid arch,,1,1. muscular process of the arytenoid cartilage,,1,2
Transversus arytenoid,1. the muscular process of one arytenoid cartilage,,1,1. the muscular process of the contralateral arytenoid cartilage ,,1,2
Vocalis,1. internal and inferior and anteromedial aspect of the thyroid cartilage,,1,1. the vocal process of the arytenoid cartilage (running along the entire length of vocal ligament),,1,2
Thyroarytenoid,1. internal and inferior and anteromedial aspect of the thyroid cartilage (lateral to vocalis),,1,1. muscular process and lateral surface of arytenoid cartilage (running alongside and lateral to the vocalis),,1,2
Oblique arytenoid,1. muscular process of an arytenoid cartilage,,1,1. muscular process of the contralaeral arytenoid cartliage,,2,3
,,,,2. fibers of the aryepiglottic muscle,,,
Aryepiglottic,1. fibers of the oblique arytenoid muscle,,1,1. lateral border of epiglottis,,1,2
Thyroepiglottic,1. fibers of the thyroarytennoid in aryepiglottic folds,,1,1. epiglottis,,1,2
External intercostals,1. inferior border of an upper rib,,1,1. superior border of a rib below (each muscle fiber runs obliquely and inserts toward the costal cartilage),,1,2
Internal intercostals,1. superior border of a lower rib (runs opposite of external intercostals,,1,1. inferior border of a rib above (each muscle fiber runs obliquely and inserts toward the costalt cartilage),,1,2
Innermost intercostals,1. superior border of a lower rib (often not well developed),,1,1. inferior border of a rib above (each muscle fiber runs obliquely and inserts toward the costalt cartilage),,1,2
Subcostals,1. inner surface of each rib near its angle,,1,1. medially on the 2nd 3rd rib below,,1,2
Diaphragm,1. sternal portion: inner xiphoid process,,9,1. central tendon of the diaphragm,,1,10
,2. costal portion: inner surface of the lower 6 ribs,6 origins,,,,,
,3. lumbar portion: upper 2 or 3 lumbar vertebrae via 2 cura,2 origins,,,,,
External oblique,1. lower borders of the lower 8 ribs,8 origins,8,1. outer lip of the iliac crest,,3,11
,,,,2. inguinal ligament,,,
,,,,3. anterior layer of the rectus sheath,,,
Internal oblique,1. middle of the iliac crest,,3,1. linea alba,,4,7
,2. lateral 1/3 of the inguinal ligament,,,2. lower borders of the lower 3 or 4 ribs,3 origins,,
,3. thoracolumbar fascia,,,,,,
Transversus abdominis ,1. inner lip of iliac crest,,9,1. linea alba,,1,10
,2. lateral 1/3 of the inguinal ligament,,,,,,
,3. thoracolumbar fascia,,,,,,
,4. cartilage of the lower 6 ribs,6 origins,,,,,
Rectus abdominus,1. cartilages of ribs 5-7,3 origins,4,1. pubic crest between pubic tubercle and pubic symphysis,,1,5
,2. xiphoid process ,,,,,,
Pyramidalis,1. ventral surface of pubis,,1,1. linea alba (midway between umbilicus and pubis,,1,2
,,,,,,,
Quadratus lumborum,1. lateral lip of iliac crest,,2,1. posteriorand inferior aspect of 12th rib,,5,7
,2. iliolumbar ligament,,,2. transverse processes of L1-L4,4 insertions,,
\end{filecontents*}
\csvreader[head to column names]{grade.csv}{}%
{\\\Muscle & \Origin & \NotesA & \numOfOriginsTotal & \Insertions & \NotesB & \numOfInsertionsTotal & \HyperedgeDegree} %
\\\hline
\end{longtable}
\endgroup

}



\end{document}